\documentclass{aa}  

\usepackage{graphicx}

\usepackage{txfonts}
\usepackage{booktabs}
\usepackage{tablefootnote}
\usepackage{stfloats}
\usepackage{multicol}
\usepackage[dvipsnames]{xcolor}

\usepackage{txfonts}

\usepackage{afterpage}
\usepackage[section]{placeins}
\usepackage{rotating}
\usepackage[T1]{fontenc}
\usepackage{amsmath}
\usepackage{textgreek}

\usepackage{subcaption}
\usepackage{silence}
\usepackage{natbib}
\bibpunct{(}{)}{;}{a}{}{,}
\usepackage{hyperref} 
\hypersetup{
    colorlinks=true,
    linkcolor=blue,
    citecolor=blue,
    filecolor=blue,
    urlcolor=blue,
}

\begin{document}

   \title{Dynamics of Y Dwarf Atmospheres}

   \author{
            C. Ak{\i}n\thanks{Corresponding author} \inst{1, 2}
            \and
            E. K. H. Lee \inst{1}
            \and
            L. Gkouvelis\inst{2}
            \and
            K. Heng\inst{2, 3,  4, 5}
          }
   \institute{
        Center for Space and Habitability, University of Bern, Gesellschaftsstrasse 6, CH-3012 Bern, Switzerland
        \and
        Faculty of Physics, Ludwig Maximilian University, Scheinerstraße 1, D-81679 München, Deutschland
        \and 
        ARTORG Center for Biomedical Engineering Research, University of Bern, Murtenstrasse 50, CH-3008, Bern, Switzerland
        \and
        University College London, Department of Physics \& Astronomy, Gower St, London, WC1E 6BT, United Kingdom
        \and
        Astronomy \& Astrophysics Group, Department of Physics, University of Warwick, Coventry CV4 7AL, United Kingdom
        }
   \date{Received 26 August 2025 / Accepted 02 February 2026}
 
  \abstract
   {The global circulation regime of the coolest class of brown dwarfs, known as the Y dwarfs, remains largely unexplored.}
   {We aim to investigate the interplay between convection, rotation, and cloud thermal feedback through a selected sample of Y dwarf atmospheric models. We explore a range of effective temperatures $400~\mathrm{K} \leq T_{\mathrm{eff}} \leq 600~\mathrm{K}$ and rotation rates $P_{\mathrm{rot}} = 2.5 \text{--} 20\ \mathrm{h}$. In this temperature range, salt and sulfide condensates are expected to form. We include $\mathrm{KCl,~Na_{2}S}$ and $\mathrm{MnS}$ clouds in our simulations to study their effect on the atmosphere. Our goal is to identify circulation regimes and emergent trends across this space, providing insights into the dynamical processes governing Y dwarf atmospheres.}
   {We run a suite of twelve general circulation models (GCMs) across the outlined parameter grid. For this purpose, we develop additional physics modules for the THOR GCM to model brown dwarf atmospheres. The THOR dynamical core is coupled to modules for interior thermal perturbations near the radiative–convective boundary, a mixing-length convection scheme, a gray two-stream radiative transfer module using Rosseland-mean opacities, and simple cloud tracers with thermal feedback and scattering.}
   {Across all simulations, the circulation resides in a radiative-forcing-dominated regime with weak winds, minimal horizontal temperature contrasts, and no persistent jets. Convection controls vertical mixing and sets the extent of the salt and sulfide cloud layers that form below the photosphere. Thermal structures equilibrate quickly and cloud radiative feedback remains insignificant, with limited variability.}
   {Y dwarf atmospheres in this parameter range, within the gray radiative transfer framework adopted in this work, remain controlled by thermal radiation from the interior, with small variability primarily set by rotation and clouds playing a secondary role. As our single-band approach does not capture spectral windows that could probe deeper into the cloud-rich layers, our constraints on cloud radiative feedback are likely conservative, and we outline possible causes and pathways toward more active regimes.}

   \keywords{
             methods: analytical -- 
             methods: numerical  -- 
             planets and satellites: atmospheres --
             planets and satellites: dynamical evolution and stability
               }

   \maketitle

\section{Introduction}
\label{sec:introduction}
Brown dwarfs are substellar objects that cannot sustain stable fusion and steadily cool over time \citep{burrows_1997, baraffe_2003, saumon_marley_2008}. Y dwarfs represent the coolest end of the brown-dwarf sequence, with currently known effective temperatures of  $250~\mathrm{K} \leq T_{\mathrm{eff}} \leq 500~\rm K$ \citep{leggett_2021} and rapid rotation periods of a few hours \citep{tannock_2021}, occupying a physical regime that overlaps with the atmospheric conditions of giant planets \citet{showman_tan_parmentier_2020}. More than ten years after Wide-field Infrared Survey Explorer (WISE) discovered the first Y dwarfs \citep{cushing_2011, kirkpatrick_2011, luhman_2014}, the \textit{James Webb Space Telescope} (JWST) is both expanding that sample and is poised to detect and characterise young sub-Jovian companions. Consequently, an expanding sample of these cool brown dwarfs has become a primary focus of recent studies \citep[e.g.,][]{leggett_tremblin_2023, beiler_2024, lew_2024}. Their atmospheric conditions make them valuable laboratories for testing dynamical and cloud physics, which could deliver insights into colder planetary atmospheres.

The spectra of Y dwarfs indicate the presence of water \citep{burrows_2003, morley_2014a}, sulfide and salt clouds \citep{morley_2012, morley_2014b_letter}, and point to vigorous vertical mixing \citep{robinson_marley_2014}, signalling that their atmospheres depart from simple equilibrium expectations \citep{lacy_burrows_2023}. Using the \citet{ackerman_marley_2001} model, \citet{morley_2012} suggested that, compared to the silicate clouds of L dwarfs, cooler brown dwarf atmospheres are expected to contain different condensates (such as $\mathrm{Cr}, \mathrm{Na_{2}S}, \mathrm{MnS}, \mathrm{KCl}$) and ran a suite of models taking these species into account, finding that cloudy models explain the spectra of these objects better than cloud-free models. \citet{morley_2014b_letter} highlight that, although photometric variability has been a focus of L/T transition studies, late-T to early Y dwarfs $(T_{\mathrm{eff}} \geq 375~K)$ also display a degree of variability \citep{cushing_2016}. They suggest that this variability could be a result of patchy cloud cover or the global circulation creating an irregular temperature structure, and mimic these effects by combining multiple 1D models. They conclude that both of these effects induce variability in different spectral ranges.

Analytical studies have provided a first guide to the relevant dynamics for these substellar atmospheres and a growing body of theory links brown-dwarf variability to interacting thermal, dynamical, cloud, and chemical processes. \citet{showman_kaspi_2013} presented a simple analytical framework, estimating the horizontal temperature variations, together with zonal- and vertical wind magnitudes, connecting them to a dimensionless efficiency parameter $\eta$, quantifying how much of the present heat flux is used to drive the circulation in these atmospheres. Using a two-layer shallow-water model that is forced from the interior by randomly varying perturbations, \citet{zhang_showman_2014} demonstrated that brown-dwarf atmospheres can settle into either of two broad circulation regimes: large-scale zonal jets or fields of local vortices. Their outcomes depended on the balance between convective forcing and radiative damping. \citet{hammond_2023_shallow_water} used a similar one-layer shallow-water model and by exploring a large parameter space covering possible brown dwarf configurations, identified four dynamical regimes. Both works conclude that, given the radiative forcing expected of Y dwarfs, they should lie in a regime dominated by zonal jets.

In addition to these analytical works, recent years saw multiple works that ran 3D general circulation models (GCMs) of field brown dwarf atmospheres. The first such work by \citet{showman_kaspi_2013} ran an idealized brown dwarf case with the MITgcm \citep{adcroft_2004_MITgcm} and found brown dwarf atmospheres to be rotation-dominated. By comparing a fast and a slow rotator case, along with analytical considerations, they concluded that, on the large scale, the flow should align in columns parallel to the rotation axis and have latitudinal variations resulting in pole-to-equator temperature differences. \citet{showman_tan_zhang_2019} performed idealized GCM simulations, forcing the atmosphere through a physically motivated parametric model of interior thermal perturbations, and they found that their models produced zonally-banded structures ubiquitously, along with oscillations in the predicted structures that they tie to the observed variability.

\citet{tan_2021a} explored the effects of cloud thermal feedback and its potential to drive atmospheric circulation in brown dwarfs using a model assuming a Cartesian geometry. They showed that cloud and rotation coupling can produce significant temperature contrasts, strong winds, and persistent, time-variable structures under suitable conditions. By coupling the MITgcm to a cloud model, \citet{tan_2021b} conducted a high-resolution study of a multitude of brown dwarf scenarios of varying rotation rates. They confirmed that cloud radiative feedback can be the main driver of dynamical activity in brown dwarf atmospheres and explored possible reasons behind the observed light curve variability \citep{cushing_2016, artigau_2018_variability}. They concluded that variability may arise both from wave activity triggered by cloud feedback and from changes in cloud coverage with viewing geometry. \citet{tan_2022} extended the model to explore jet formation and tracer transport in brown dwarf and isolated giant planet atmospheres. The study found that thermal forcing can generate strong zonal jets, vertical shear, and long-term oscillations in equatorial winds, depending on the strength of radiative and frictional damping. It also showed that these circulations can mix cloud particles and chemical tracers vertically, with the extent of mixing varying by particle size and chemical properties.

In a series of works, \citet{lee_2023_BD} and \citet{lee_2024_BD_clouds}, using ExoFMS \citep{lin_2004_FMS, lee_2020_simple_BD}, presented GCM simulations of brown dwarf scenarios coupling their radiative transfer, chemistry and clouds with increasingly more sophisticated treatments of these physical processes. They found that these processes feed back onto each other in a complex fashion and are crucial in explaining the observed dynamical and thermal structure, along with the variability of a given brown dwarf atmosphere.

In a related work, \citet{lee_ohno_2025} developed a two-moment cloud microphysics module, coupled it to the ExoFMS GCM and applied it to an isolated Y dwarf case. They found that the parameter regime they tested produced weak dynamical activity and emphasized the need for a wider sweep of parameter space, an issue we address with the current work. This work extends GCM studies of cold brown dwarf atmospheres with a targeted survey of Y dwarf parameter space using a unified physical treatment. We span rotation rates and effective temperatures characteristic of early Y dwarfs and exclude the cold tail where $\mathrm{H_{2}O}$ clouds form, while variations in $R_p$, $\log g$, metallicity, and disequilibrium chemistry are left for future studies. 

The structure of the paper is organized as follows. In Section \ref{sec:characteristic_flow_quantities}, we introduce diagnostic quantities that have been used in previous theoretical works on brown dwarfs and give an analytical order-of-magnitude estimate of the dynamical regime we expect Y dwarfs to fall into. Section \ref{sec:run_cases} gives an account of our chosen sample of models, along with our reasoning for the presented modelling choices. Sections \ref{sec:dynamical_core}-\ref{sec:thermal_feedback_and_scattering_of_clouds} detail our GCM setup along with the additional physics modules that were developed for this study. Section \ref{sec:results} goes over our main results, detailing identified trends in our sample of 12 Y dwarf cases, their corresponding flow regimes and the influence of mineral clouds in their atmospheres. Subsequently, in Section \ref{sec:discussion}, we discuss how our results fit into the existing literature, offer insights from the first large-scale GCM study on Y dwarfs, and lastly, highlight possible improvements for future studies.

\section{Methodology}
\subsection{Characteristic flow quantities}
\label{sec:characteristic_flow_quantities}

To ground our global circulation modelling in first-principles expectations and as a prelude to the numerical simulations, we derive order-of-magnitude estimates for the thermal and dynamical regimes utilizing prior shallow water theoretical work done on brown dwarf atmospheres. The analytical framework of \citet{hammond_2023_shallow_water} uses a dimensionless two-parameter space to describe the circulation regimes of a brown dwarf atmosphere. The first characteristic quantity is the thermal Rossby number, 
\begin{equation}
    Ro_{\mathrm{T}} = \frac{\Phi_{0}}{\left(2\Omega R_{\mathrm{p}}\right)^{2}}, 
\end{equation}
where $\Phi_{0}$ is the equilibrium geopotential. This can be connected to the standard Rossby number \citep{holton_1992, showman_guillot_2002}, defined as,
\begin{equation}
    Ro = \frac{U}{f L},
\end{equation}
through thermal wind balance $\Phi_{0} = 2 \Omega U R_{\mathrm{p}}$ \citep{mitchell_vallis_2010}. Here, $U~ [\mathrm{m~s^{-1}}]$ is the characteristic wind speed, $L~[\mathrm{m}]$ the characteristic length scale, and $f=2 \Omega \sin{\theta}~[\mathrm{s^{-1}}]$ the Coriolis parameter at latitude $\theta$ for a planet rotating at rate $\Omega~[\mathrm{rad~s^{-1}}]$. While, for the purposes of this section, we can estimate $Ro_{\mathrm{T}}$ via an idealized thermal wind calculation, within the GCM we calculate the geopotential as \citep{zhang_showman_2014},
\begin{equation}
\label{eq:geopotential}
    \Phi_{0} = \gamma~\frac{R_{d}^{2}}{c_{p}}~T,
\end{equation}
where $R_{d}~[\mathrm{J~K^{-1}~kg^{-1}}]$ is the specific gas constant, $c_{p}~[\mathrm{J~K^{-1}~kg^{-1}}]$ the specific heat at constant pressure, $T~[\mathrm{K}]$ the temperature and $\gamma$ a measure of adiabacity given by,

\begin{equation}
\label{eq:adiabacity_measure}
    \gamma = \left[1 +  \left(\frac{c_{\mathrm{p}}}{g}\frac{dT}{dz }\right) \right].
\end{equation}

Both definitions of the Rossby number gauge the relative importance of inertial to Coriolis forces, where small $\left(\mathrm{Ro} \ll 1\right)$ values imply rotation‑dominated dynamics. The second coordinate is the dimensionless radiative timescale,
\begin{equation}
\label{eq:normalized_tau_rad}
    \hat{\tau}_{\mathrm{rad}} = 2\Omega \tau_{\mathrm{rad}},
\end{equation}
which incorporates the physical radiative relaxation time \citep{showman_guillot_2002},
\begin{equation}
    \tau_{rad} \sim \frac{p}{g} \frac{c_{p}}{\sigma_{\mathrm{SB}} T^{3}},
\end{equation}
where $p~[\mathrm{Pa}]$ is pressure, $g~[\mathrm{m~s^{-2}}]$ surface gravity and $\sigma_{\mathrm{SB}}~[\mathrm{W~m^{-2}~K^4}]$ the Stefan–Boltzmann constant. Although different physical conditions can lead to the same pair of parameter values $(Ro_{\mathrm{T}},\hat{t}_{\mathrm{rad}})$, \citet{hammond_2023_shallow_water} state that every point in this plane corresponds to a unique qualitative circulation state, providing a useful description of the underlying dynamical state of the atmosphere.

Unfortunately, to estimate the Rossby number and dimensionless radiative timescale requires specifying the values of quantities that are the outcomes of GCM simulations. We thus use analytical scaling relations to make educated guesses for the values of these quantities. To estimate the horizontal wind velocity $U$ we adopt the wave‑driven scaling relation of \citet{showman_kaspi_2013}. By combining geostrophic thermal-wind balance with a simple radiative-damping closure, they derived an expression of the characteristic horizontal wind speed $U$ in a fast-rotating, non-irradiated atmosphere as,
\begin{equation}
U \simeq\sqrt{\tfrac{\eta c_p T}{4}} ~\frac{\ell\,\Delta \tilde z\,HN}{f},\qquad    
\end{equation}
where $H~[\mathrm{m}]$ is the pressure scale height, defined as 
\begin{equation}
    H = \frac{k_{B}T}{m g},
\end{equation}
and  $N~[\mathrm{s^{-1}}]$is  the Brunt–Väisälä frequency,
\begin{equation}
    N = \sqrt{\frac{g}{T}  \left(\frac{dT}{dz} + \frac{g}{c_{p}} \right)},
\end{equation}
$\ell~[m^{-1}]$ the characteristic wavenumber, $\Delta\tilde z~[H]$ the vertical thickness of the overturning cell in scale heights, and $\eta$ the fraction of the interior heat flux that goes into driving atmospheric waves, called the dimensionless wave driving efficiency. 

Although the Y dwarf parameter range spans $250~\mathrm{K} \leq T_{\mathrm{eff}} \leq 500 ~\mathrm{K}$, we assume $400~\mathrm{K}\leq T_{\mathrm{eff}} \leq 600~\mathrm{K}$ for our calculations presented in this section, matching our modelled cases detailed in Section \ref{sec:run_cases}. The radius is taken to be equal to the radius of Jupiter $R_{\mathrm p}=R_{\mathrm J}$, since the radii of brown dwarfs are not expected to change significantly past their initial evolution and typically fall very close within one Jupiter radius \citep{burrows_1997, legget_2017}. For the surface gravity, we choose a representative value of $\log g = 4.5$ and later discuss the qualitative impact of changing this parameter. Rotation periods in the range $P_{\mathrm{rot}} = 2.5 \text{--} 20\ \mathrm{h}$ encompass most of the current observations \citep{legget_2017, tannock_2021} and imply angular velocities $\Omega = 9 \times 10^{-5} \text{--} 7 \times 10^{-4}~\mathrm{rad~s^{-1}}$. We evaluate the Coriolis parameter at a representative mid–latitude $\theta = 45^{\circ}$. Assuming an approximate solar–composition mixture, with volume mixing ratios (VMRs) $\chi_{\mathrm{H_{2}}} = 0.84$ and $\chi_{\mathrm{He}} = 0.16$, yields a mean molecular weight $\mu = 2.33$, provides the thermodynamic background, with a specific gas constant $R_{d} = \frac{k_{\mathrm{B}}}{\mu m_{u}} = 3569~\mathrm{J~K^{-1}~kg^{-1}}$ and specific heat at constant pressure $c_p = \frac{R_d}{\kappa_{\mathrm{ad}}} = 12039~\mathrm{J~K^{-1}~kg^{-1}}$, where $m_{u}$ is the atomic mass unit and $\kappa_{\mathrm{ad}}$ is the adiabatic gradient. 

Following \citet{showman_kaspi_2013} and aiming for simplicity, we assume an isothermal atmosphere. This allows us to calculate $ N H = R_{d} \sqrt{\frac{T}{c_{p}}}$. For wave driving, we adopt the characteristic horizontal wavenumber of the same work $\ell = 3 \times 10^{-7}~\rm  m^{-1}$, which corresponds to a characteristic length scale of $\approx 10^{4}~\mathrm{km}$, matching observational values \citep{simon_miller_2006} for Jupiter. A quick comparison to the Rossby deformation radius, as it is defined in \citet{showman_kaspi_2013},
\begin{equation}
    L_{D} = \frac{\Delta \tilde{z} H N}{f},
\end{equation}
yields a similar average value of $\approx 5 \times 10^{3}~\mathrm{km}$ when calculated over all the ranges of values given above. The vertical extent of the overturning cell is assumed to range between $\Delta \tilde z = 1 - 3$ \citep{showman_kaspi_2013}, meaning that the circulation spans one to three scale heights.
Following the same study, we define the wave-driving efficiency as the fraction of the interior heat flux that goes into accelerating the mean flow,
\begin{equation}
    \eta \equiv \frac{\mathcal{A}\,\Delta u\,p}{F_{\rm OLR}\,g},
\end{equation}
where $\mathcal{A}$ is the eddy acceleration, $\Delta u$ a characteristic wind speed and $F_{\rm OLR}$ the emitted flux. Using observationally inferred $\mathcal{A}$, $\Delta u$, $p$, $F_{\rm OLR}$ for Earth and Jupiter, \citet{showman_kaspi_2013} infer $\eta \sim 10^{-3}$ for both cases and suggest a range of  $\eta \sim 10^{-4}\text{--}10^{-2}$ for brown dwarf studies. In our calculations we explore values over $\eta = 10^{-4}$ to $10^{-1}$, a range that encompasses very weak to vigorous wave forcing.

\begin{figure}
    \centering
    \includegraphics[width=\columnwidth]{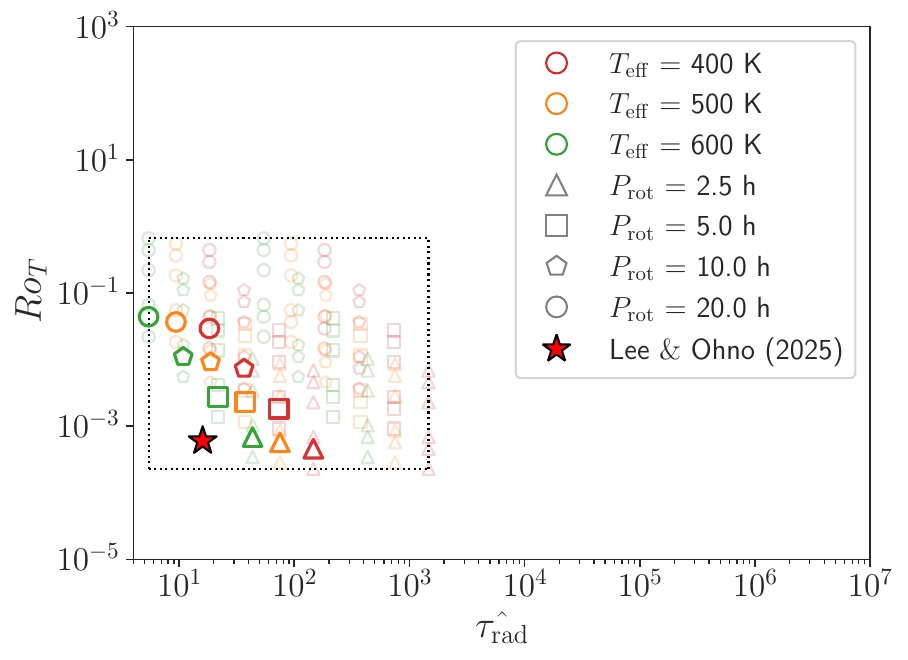}
    \caption{A summary of our calculations, presented in Section \ref{sec:characteristic_flow_quantities}. Highlighted are the cases with gravity $\log g = 4.5$, a wave-driving efficiency of $\eta=10^{-3}$ and a vertical scale of $\Delta \tilde{z} = 2$ at the approximate photosphere level $p_{\mathrm{photosphere}} = 0.1 \rm~bar$. Other possible combinations are overplotted in fainter colors to illustrate the complete space spanned by the range of values considered in this study. Dotted lines mark the edges of minimum and maximum values. The red star corresponds to ($T_{\mathrm{eff}} = 400~\mathrm{K}~;~ P_{\mathrm{rot}} = 10~ \rm h$; $\log g = 4.46$) and is from a recent GCM study, included as a reference point.}
    \label{fig:expected_regimes}
\end{figure}

We consider two photospheric pressure values $p = [0.1, 1]~\mathrm{bar}$ in the radiative timescale expression. Combining these values with the rotation periods above gives a dimensionless radiative time $\hat{\tau}_{\mathrm{rad}}$ that lies between $5$ and $10^{3}$. The resulting pair $(Ro_{\mathrm{T}}, \hat{\tau}_{\mathrm{rad}})$ therefore situates Y–dwarf atmospheres in the circulation regime characterized by radiative forcing with no jets expected to form \citep{hammond_2023_shallow_water}. The full range of possible values is illustrated in Figure \ref{fig:expected_regimes}, with the extent of the axes corresponding to the ones presented in \citet{hammond_2023_shallow_water} to enable a direct comparison.

Adjusting each parameter causes a corresponding shift in the position on the plot. Lower effective gravities of $\log(g) = 3.5~\text{--}~4$ shift the entire plot to the right and higher effective gravities shift it to the left. Although surface gravity affects both the predicted thermal Rossby numbers and radiative timescales, in our idealized framework, its effect is more dominant on the radiative timescales. Selecting a deeper photosphere, e.g., $p= 1~ \rm bar$ (also illustrated with fainter points in Figure \ref{fig:expected_regimes}), has the same effect of shifting the points to the right and different wave-driving efficiencies shift the plot upwards or downwards, as expected.

\subsection{Design of simulation cases}
\label{sec:run_cases}

\begin{figure}
    \centering
    \includegraphics[width=\columnwidth]{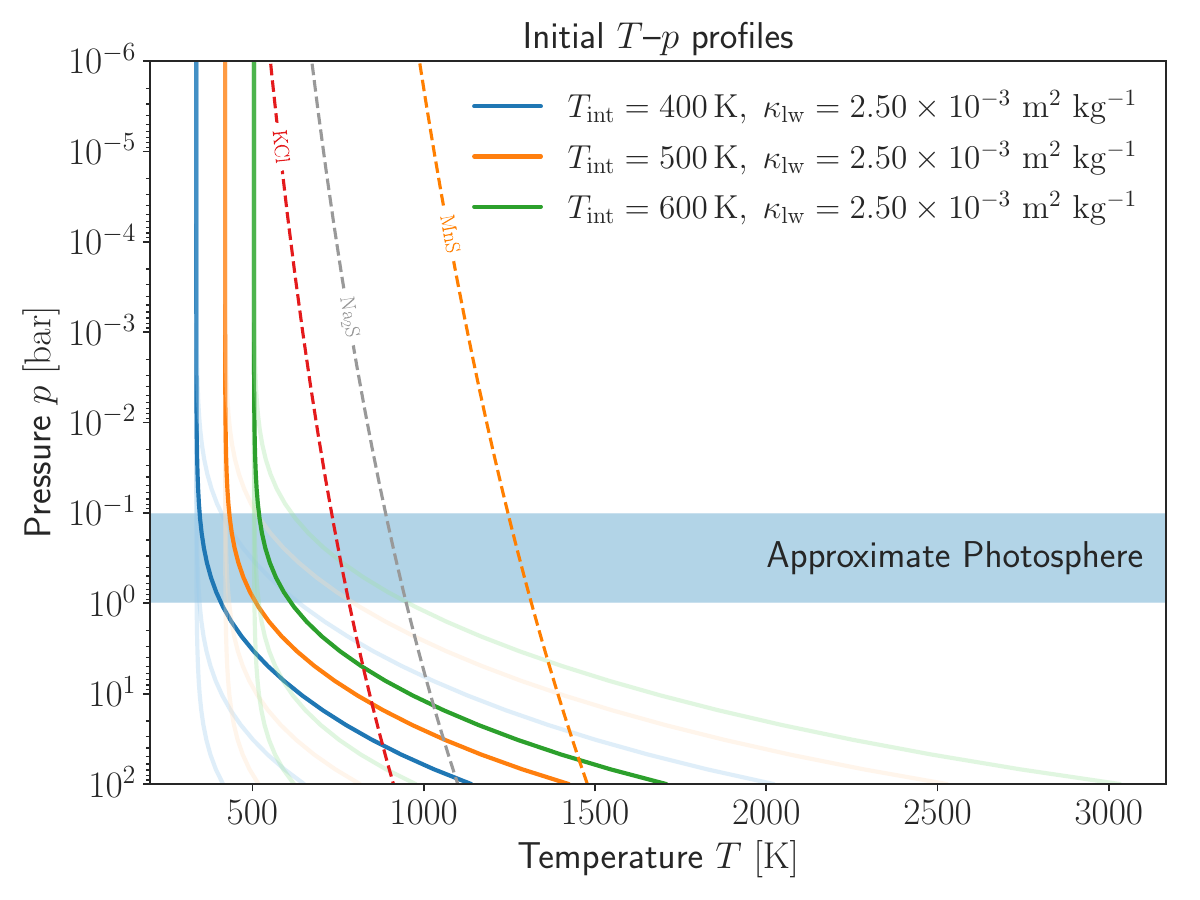}
    \caption{Various $T\text{--}p$ profiles overplotted to illustrate the initial conditions from which the simulations were spun up. The solid lines correspond to the $\kappa_{\mathrm{lw}} = 2.5 \times 10^{-3} ~\rm m^{2}~kg^{-1}$ curves with the lighter shades of each color illustration curves corresponding to the same internal temperature $T_{\mathrm{int}}$ with varying long-wave opacities of $\kappa_{\mathrm{lw}} = [2.5 \times 10^{-5}, 2.5 \times 10^{-4}, 2.5 \times 10^{-2}] \rm ~m^{2}~kg^{-1}$.  Also overplotted are condensation curves for the cloud condensate species $\rm KCl,~MnS,~Na_2S$. The shaded horizontal region ($p = 0.1~\text{--}~1~\rm bar$) corresponds to the approximate photosphere.} 
    \label{fig:eddington_profs}
\end{figure}

To capture the centre of the temperature distribution of the early Y dwarf regime, while avoiding the parameter space where $\mathrm{H_{2}O}$ clouds would form, we adopt $400~\mathrm{K}\leq T_{\mathrm{eff}} \leq 600~\mathrm{K}$, where the warm end ($T_{\mathrm{eff}} = 600~\mathrm{K}$) overlaps with the late-T regime. We chose our representative Y dwarf scenarios to have a radius of $R_{\mathrm{jup}}$, together with an effective surface gravity of $\log g = 4.5$. Although observed Y dwarf radii vary about $25~\%$, they are mostly around one Jupiter radius \citep{burrows_1997, legget_2017}. For the surface gravity, while retrieval \citep[e.g.,][]{zalesky_uniform_2019} and modelling \citep[e.g.,][]{lacy_burrows_2023} studies suggest a range of $3.5 \leq \log(g) \leq 5$, we select a single representative value of $\log g = 4.5$ . As discussed in Section \ref{sec:characteristic_flow_quantities}, changing $g$ mainly produces a horizontal shift in $\hat{\tau}_{\mathrm{rad}}$ on Figure \ref{fig:expected_regimes}, whereas the shallow-water regimes identified by \citet{hammond_2023_shallow_water} vary most strongly in the vertical $(Ro_{\mathrm{T}})$ direction. We, therefore, keep the surface gravity fixed and concentrate our computational efforts on exploring a wider range of effective temperatures, rotation rates and cloud species. To decide which cloud species to include in our simulations, we look at previous studies \citep{morley_2012, gao_benneke_2018_clouds} and remark that for the temperature range we find ourselves in, the main relevant cloud species are $\mathrm{KCl,~MnS}$ and $\mathrm{Na_{2}S}$. Figure \ref{fig:eddington_profs} illustrates the condensation curves for these species overplotted with our initial $T\text{--}p$ profiles. Finally, we assume solar metallicity for all of our runs. Current observations put most Y dwarfs close to solar metallicities \citep{legget_2017} and since most of these objects are observed nearby our Solar neighborhood, this provides a reasonable assumption. Our constructed suite of run cases that are briefly summarized in Table \ref{tab:run_parameters}.

We start our GCM runs from rest with a gray analytical $T\text{--}p$ profile \citep{milne_1921, eddington_1926} derived for a self-luminous object in the gray opacity limit,
\begin{equation}
\label{eq:eddington_tp}
    T^{4} = \frac{3 T_{\mathrm{int}}^{4}}{4} \left[ \frac{2}{3} + \tau_{\mathrm{lw}}\right],
\end{equation}
with $T$ describing the temperature structure with optical depth, $T_{\mathrm{int}}$ the bottom boundary internal temperature of the atmosphere, and $\tau_{\mathrm{lw}}$ the long-wave optical depth. The optical depth term $\tau_{\mathrm{lw}}(p)$ controls the location of the photosphere and is solely used to construct the initial $T\text{--}p$ profile. Following the radiative equilibrium solutions calculated in previous studies for brown dwarf atmospheres \citep[e.g.,][]{burrows_2006}, we aim for an adiabatic interior, an isothermal upper atmosphere and a profile with the approximate photosphere ($p \approx 0.1~\text{--}~1~\rm bar$) corresponding to the effective temperature for each case. Therefore, we choose a longwave gray opacity of $\kappa_{\mathrm{lw}} = 2.5 \times 10^{-3} ~\rm m^{2}~kg^{-1}$ to generate our initial $T\text{--}p$ profile using the relation \citep{deitrick_thor2.0_2020}
\begin{equation}
    \tau_{\mathrm{lw}}  = \tau_{0}  \frac{p}{p_{\mathrm{ref}}} + (\frac{1}{f_{\mathrm{lw}}}- 1) ~\tau_{0} ~\left( \frac{p}{p_{\mathrm{ref}}} \right)^{2}
\end{equation}
where $f_{\mathrm{lw}}$ is the strength of the uniformly-mixed absorbers set to $0.5$,  $\tau_{0}=\kappa_{\mathrm{lw}} \frac{p_{\mathrm{ref}}}{g}$ the reference optical depth, $p_{\mathrm{ref}}$ the reference pressure of the bottom boundary set to $100~\mathrm{bar}$ for our current set of simulations. It should be noted that the quadratic dependence $\tau_{\mathrm{lw}} \propto p^2$ makes this solution convectively unstable at initialization. To remove this initial instability without altering the target profile, we apply a brief initialization (10 computational steps of $100~\mathrm{s}$) with dry convective adjustment and then continue with our mixing-length theory (see Section \ref{sec:mixing_length_theory}) routine. The $T\text{--}p$ profile after this step is essentially unchanged and the simulation proceeds stably. Figure \ref{fig:eddington_profs} illustrates various examples for initial $T \text{--}p$ profiles, and we can see that the $\kappa_{\mathrm{lw}} = 2.5 \times 10^{-3}~\rm m^{2}~kg^{-1}$ case fits the characteristics we were aiming for.

In our tests, we also initialized simulations using an isothermal temperature profile, $T \left(p\right) = T_{\mathrm{int}}$, letting the dynamics dictate the evolution of the $T\text{--}p$ profile.  This approach ultimately converged to the same steady-state $T\text{--}p$ structure as with the gray profile given by Eq. \ref{eq:eddington_tp}. However, the latter profile notably accelerates convergence to the steady state, as it closely approximates the final equilibrium solution. This approach also circumvents possible numerical issues where starting from an isothermal profile can generate a confined cloud condensate layer at the bottom of the atmosphere that results in a physically unrealistic scenario.

\begin{table}
    \centering
    \small
    \renewcommand{\arraystretch}{1.2}
    \caption{Planetary and atmospheric parameters defining our suite of simulations.}
    \label{tab:run_parameters}
    \resizebox{\columnwidth}{!}{
    \begin{tabular}{@{}l c l r@{}}
        \toprule
        \textbf{Symbol} & \textbf{Descriptor} & \textbf{Unit} & \textbf{Value} \\ 
        \midrule
        $T_{\mathrm{eff}}$ & Effective Temperature & [$\mathrm{K}$] 
          & 400,\;500,\;600 \\
        
        $P$ & Rotation Period & [$\mathrm{h}$] 
          & 20,\;10,\;5,\;2.5 \\
        
        $R_{p}$ & Radius & [$R_{\mathrm{jup}}$] 
          & 1.0 \\
          
        $\log g$ & Gravity & [-] 
          & 4.5 \\
        $R_{d}$  & Specific gas constant & [$\mathrm{J}~\mathrm{K^{-1}}~\mathrm{kg^{-1}}$] 
          & 3569 \\
        $c_{p}$  & Specific heat capacity & [$\mathrm{J}~\mathrm{K^{-1}}~\mathrm{kg^{-1}}$] 
          & 12039 \\
        $\log_{10}\left([M/H]\right)$ & Metallicity & [-] 
          & 0.0 \\
        
        $\rm Cloud$ & Cloud Species & [-] 
          & $\rm KCl,\;MnS,\;Na_2S$ \\
        
        $t_{\mathrm{run}}$ & Run Time & [Earth days] 
          & 1000 \\
        \bottomrule
    \end{tabular}
    }
\end{table}

Recent studies by \citet{tan_2021b}, \citet{lee_2023_BD} and \citet{lee_2024_BD_clouds} suggest that these atmospheres reach a statistical steady state after a couple of hundred days. Our simulations confirm this fact and we choose a uniform simulation time of 1000 Earth days for all of our presented results, aiming for a reasonable yet computationally feasible value. The last critical consideration for the general structure of runs is the average horizontal resolution. Since Y dwarfs are rapid rotators \citep{legget_2017, tannock_2021}, we require a high enough resolution to capture the length scale of the dominant dynamical features. To estimate the needed length scale, we take the minimum possible Rossby deformation radius $L_{\mathrm{D}}$ for our setup, calculated for a range of plausible parameters (as they are presented in Section \ref{sec:characteristic_flow_quantities}). The minimum, achieved only for the fastest rotators ($P_{\mathrm{rot}}=2.5~\mathrm{h}$), is $L_{\mathrm{D}}\approx 6.5\times10^{5}~\mathrm{m}$, corresponding to a refinement level $g_{\mathrm{level}} \approx 7$ as described in \citet{mendonca_thor_2016,deitrick_thor_2022}. At the opposite end, the slowest rotators ($P_{\mathrm{rot}}=20~\mathrm{h}$) have a larger minimum $L_{\mathrm{D}}\approx 5.271\times10^{6}~\mathrm{m}$, which would only require $g_{\mathrm{level}}\approx 4$. To conservatively resolve small-scale structure and to ease cross-comparison, while also aiming to remain computationally feasible, we adopt $g_{\mathrm{level}}=7$ for all but the slowest cases. For the $P_{\mathrm{rot}}=20~\mathrm{h}$ runs, we use $g_{\mathrm{level}}=6$, which still comfortably resolves the deformation scale at those rotation rates. These considerations land us at a chosen average angular resolution of $\bar{\theta} \approx 1^{\circ}$ for the slowest rotators and $\bar{\theta} \approx 0.5^{\circ}$ for the rest of the runs and are comparable to the high-resolution simulations performed by \citet{tan_2021a} and \citet{tan_2022}.

\begin{table}
    \centering
    \Large
    \renewcommand{\arraystretch}{1.3}
    \caption{Secondary parameters governing clouds, convection, and numerics.  \protect\footnotemark}
    \label{tab:cloud_numerics}
    \resizebox{\columnwidth}{!}{
    \begin{tabular}{@{}l c l r@{}}
        \toprule
        \textbf{Symbol} & \textbf{Descriptor} & \textbf{Unit} & \textbf{Value} \\
        \midrule
        \multicolumn{4}{l}{\textit{Cloud physics}}\\
        $\tau_{\mathrm{deep}}$    & Deep vapor replenishment timescale     & [s]            & $10^{3}$ \\
        $\chi_{v,\mathrm{deep}}$     & Deep vapor VMR                         & [-]            & $[1.17,\,2.63,\,8.20] \times 10^{-7}$ $\ast$ \\
        $\sigma$                  & Particle-size-distribution std.         & [-]            & 2.0 \\
        $r_{\mathrm{m}}$          & Median particle size                    & [m]            & $10^{-6}$ \\
        $\tau_{\mathrm{c}}$       & Cloud condensation timescale             & [s]            & 120 \\
        \addlinespace
        \multicolumn{4}{l}{\textit{Thermal perturbations}}\\
        $n_{\mathrm{burn}}$       & Burn-in iterations                      & [-]            & 100 \\
        $t_{\mathrm{storm}}$      & Storm timescale                         & [s]            & $10^{5}$ \\
        $T_{\mathrm{amp}}$        & Perturbation amplitude                  & [K\,s$^{-1}$]  & $[2.15,\,6.80,\,17.4] \times 10^{-5}$ \( \dag  \) \\
        $p_{\mathrm{rcb}}$        & Perturbation boundary level             & [bar]          & 10 \\
        $n_{f}$                   & Forcing wavenumber                      & [-]            & 40 \\
        \addlinespace
        \multicolumn{4}{l}{\textit{Time-stepping}}\\
        $\Delta t_{\mathrm{dyn}}$ & Dynamical time step                     & [s]            & 100 \\
        $\Delta t_{\mathrm{MLT}}$ & Mixing-length time step                 & [s]            & 0.5 \\
        \addlinespace
        \multicolumn{4}{l}{\textit{Numerical hyperparameters}}\\
        $g_{\mathrm{level}}$      & Grid-refinement level                   & [-]            & 6,\,7 \( \ddag  \) \\
        $v_{\mathrm{level}}$      & Number of vertical levels               & [-]            & 50 \\
        $D_{\mathrm{hyp,h}}$       & 6$^{\text{th}}$-order horiz.\ hyperdiffusion & [-]       & 0.001 \\
        $D_{\mathrm{hyp,v}}$       & 6$^{\text{th}}$-order vert.\ hyperdiffusion  & [-]       & 0.001  \\
        $D_{\mathrm{div}}$        & 4$^{\text{th}}$-order divergence damping    & [-]       & 0.01   \\
        $k_{\mathrm{sp}}$         & Sponge‐layer strength (horizontal/vertical)             & [s$^{-1}$]     & $10^{-3}/10^{-4}$ \\
        $\eta_{\mathrm{sp}}$      & Bottom of sponge layer ($z/z_{\text{top}}$) & [-]       & 0.8 \\
        $\eta_{\mathrm{lats}}$    & Latitude bins in sponge layer           & [-]            & 20 \\
        \bottomrule
    \end{tabular}}
\end{table}
\footnotetext{Values marked with ($\ast$) are the deep vapor reservoir VMRs for KCl, MnS, and Na$_2$S, respectively. Values marked with ($\dag$) are temperature pertubation amplitudes for $T_{\mathrm{int}}=400,\;500,\;600$ K, respectively. The grid-refinement level $g_{\mathrm{level}}$ ($\ddag$) is 6 for the $P_{\mathrm{rot}}=20$ h models and 7 for the $P_{\mathrm{rot}}=10,\;5,\;2.5$ h models.}

\subsection{The dynamical core}
\label{sec:dynamical_core}
In our study, we use the THOR\footnote{\url{https://github.com/exoclime/THOR}} dynamical core \citep{mendonca_thor_2016, mendonca_w43b_2018a, mendonca_w43b_2018b, deitrick_thor2.0_2020, deitrick_thor_2022, noti_examining_2023} to simulate the atmospheric circulation of Y dwarf atmospheres. THOR is an open-source, fully three-dimensional dynamical core that uses a horizontally explicit, vertically implicit \citep{tomita_satoh_2002, wicker_skamarock_2002} time integrator on an icosahedral grid \citep{tomita_shallow_2001} to solve the non-hydrostatic deep Euler equations. This split-time integration scheme allows THOR to maintain numerical stability with vertically propagating waves in its solutions while still taking a reasonable time step, which would otherwise be constrained by the Courant–Friedrichs–Lewy condition. Our selected grid structure avoids the usual limitations of latitude–longitude grids \citep{staniforth_thuburn_2012}, which experience singularities and resolution difficulties at the poles. Such problems arise in all non-uniform grid structures characterized by fixed points and substantial variations in cell sizes or resolution crowding. The icosahedral grid proposed by \citet{tomita_shallow_2001} and \citet{tomita_new_2004} provides a quasi-uniform horizontal grid layout. In the vertical direction, the THOR model employs an altitude grid and is implemented in two possible configurations. The first configuration, described in \citet{mendonca_thor_2016} and \citet{deitrick_thor2.0_2020}, uses a uniformly spaced altitude grid. The bottom-layer altitude is obtained from the ideal gas law, and the user defines the top of the atmosphere as a free parameter. An alternative introduced in \citet{noti_examining_2023} provides more detailed control over the vertical spacing, enabling a non-uniform vertical structure. Although the default THOR model primarily solves the non-hydrostatic deep Euler equations, as outlined in \citet{deitrick_thor2.0_2020}, \citet{deitrick_thor_2022}, and \citet{noti_examining_2023}, it also provides the option to implement various simplifications to the governing equations. These alternatives include quasi-hydrostatic deep or hydrostatic shallow formulations, each with its own assumptions. In this work, we use the non-hydrostatic deep equations with a uniformly spaced vertical grid structure.

To dissipate energy that accumulates at the grid scale, every run employs two explicit schemes, horizontal and vertical hyperdiffusion, and a divergence-damping operator, together with a Rayleigh sponge near the model top. The hyperdiffusion terms remove small-scale enstrophy, the divergence operator acts directly on the flow’s divergence field, and the sponge relaxes all wind components toward their zonal means above a prescribed height, thereby absorbing waves that would otherwise reflect off the top of the atmosphere. No basal drag is applied at the lower boundary. The numerical orders, coefficients, and sponge parameters match those used in our works using the THOR GCM \citep[e.g.,][]{akin_etal_2025_global} and are listed in Table \ref{tab:cloud_numerics}.

\subsection{Additional physics modules for brown dwarfs}
As introduced in \citet{mendonca_thor_2016} and \citet{deitrick_thor2.0_2020}, the THOR GCM was built from the ground up for exoplanetary atmospheres and has, at the time of writing, been mainly applied to model hot Jupiters \citep[e.g.,][]{mendonca_w43b_2018a, deitrick_thor_2022, noti_examining_2023, akin_etal_2025_global}. Brown dwarfs, meanwhile, occupy an intermediate region between giant planets and low-mass stars. Although there are significant similarities between brown dwarfs and giant planets, such as similar surface gravities, molecular compositions, and cloud chemistry \citep{showman_tan_parmentier_2020}, there are also distinct differences that require the development of additional physics modules to accurately model their atmospheres.

The main differences between the circulation regimes of brown dwarfs and hot Jupiters arise from the fact that hot Jupiter atmospheres are externally forced by their host star. Together with tidal locking and a lack of significant internal heating, the dynamics of these atmospheres can be characterized by a pronounced horizontal (day-night) heating gradient and the interplay of this gradient with planetary rotation. In field brown dwarfs, the situation is inverted. These objects lack any external forcing and slowly radiate away the residual heat from their formation, and in more massive cases, the negligible output of halted nuclear fusion. Their atmospheres are therefore driven primarily by the interior. The resulting thermal forcing is vertical, extending from the deep convective interior up to the observable photosphere, and, when combined with the typically rapid rotation of field brown dwarfs, sets up a circulation in which convective transport and vertical mixing play a central part. Accordingly, we implement a routine that introduces thermal perturbations at the approximate radiative-convective (RCB) boundary and we replace the present convective adjustment scheme \citep{deitrick_thor2.0_2020} with a mixing-length-theory formulation that treats sub-grid convection more realistically. Many widely used exoplanetary GCMs implement a convective adjustment scheme \citep[e.g.,][]{heng_2011b, leconte_3d_2013, deitrick_thor2.0_2020} that mixes the entropy in a convectively unstable zone, approximating convection as an instantaneous process compared to the dynamics. While \citet{lee_2023_BD} model a simplified brown dwarf case using convective adjustment, they also subsequently advocate for the improvements offered by their mixing-length theory implementation in their follow-up work \citep{lee_2024_BD_clouds}. Fully convection-resolving global models \citep[e.g.,][]{lefevre_2022_convection_BD} remain prohibitively expensive for the parameter space explored here.

Throughout this study, we confine ourselves to field brown dwarfs and therefore neglect any external instellation. This choice has a direct bearing on the radiative transfer strategy. THOR presently contains three schemes: the original two-stream double-gray module \citep{mendonca_thor_2016, mendonca_w43b_2018b, deitrick_thor2.0_2020}; a two-stream, picket-fence non-gray module modelled on the work of \citet{lee_simulating_2021} and described by \citet{noti_examining_2023}; and a multi-wavelength "improved two-stream" method (dubbed THOR+HELIOS in \citealt{deitrick_thor_2022}), that couples the THOR dynamical core to the HELIOS radiative transfer solver and provides a rigorous treatment of scattering by medium-sized and large aerosols \citep{heng_analytical_2017, heng_analytical_2018, malik_helios_2017, malik_self-luminous_2019}. For the cloud-forming, non-irradiated objects considered here, the existing options are not ideal for our goals. The picket-fence module devotes bands to stellar shortwave and offers little benefit without instellation, while the multi-wavelength THOR+HELIOS coupling is too costly for long integrations with time-varying, cell-wise cloud properties. We require an efficient radiative transfer scheme for the planetary emission that supplies a locally evaluated opacity in every grid cell so that evolving cloud opacities can be coupled to the thermal structure calculated by the solver. We therefore adopt a single thermal band with locally evaluated Rosseland-mean gas and cloud opacities, mirroring the approach of many previous brown dwarf studies \citep[e.g.][]{tan_2021b, tan_2022, lee_2023_BD}.

In summary, we have developed four complementary additions to THOR: an interior thermal perturbation model, a mixing-length sub-grid convection scheme, a gray radiative transfer module using a Rosseland-mean gray opacity scheme, and a simplified cloud tracer module that includes thermal feedback and scattering due to clouds. Our implementation follows the approaches set out by \citet{showman_tan_zhang_2019}, \citet{tan_2021a}, \citet{tan_2022} and \citet{lee_2023_BD}, and adapts them for the THOR GCM, as detailed in the following subsections. A quick overview of our parameters for these additional physics modules can be found in Table \ref{tab:cloud_numerics}.

\subsubsection{Radiative transfer}
\label{subsection:2s_nongray_IR_RT}

We extend THOR with a computationally efficient, gray radiative transfer module tailored to field brown dwarfs, adding the ability to couple cloud-opacity feedback while eliminating costly stellar-insolation calculations that are irrelevant for our non-irradiated targets. To achieve this, we prune the original picket-fence scheme \citep{noti_examining_2023}, retaining a single IR band for the planetary emission. This makes the scheme well-tested and efficient while satisfying our requirements for a temperature- and pressure-dependent local opacity structure that can couple to the cloud module we detail in Section \ref{sec:modelling_of_clouds}.

Our method follows a classic two-stream, short-characteristics formalism with linear interpolants \citep{Olson_Kunasz_1987} to solve the radiative transfer equation. At every timestep, the local gas opacity is interpolated from the \citet{freedman_2014} grid, which provides analytical fits to the Rosseland‐mean opacity $\kappa_{\mathrm{R}}$, defined as,

\begin{equation}
    \frac{1}{\kappa_{\mathrm{R}}(T,p)}=\left(\displaystyle \int_0^\infty \tfrac{\partial B_\nu}{\partial T}\,d\nu\right)^{-1} \left(\displaystyle \int_0^\infty \tfrac{1}{\kappa_\nu} \,\tfrac{\partial B_\nu}{\partial T}\,d\nu \right).
\end{equation}

The Rosseland mean weighs the frequency-dependent opacity $\kappa_\nu$ by the temperature gradient of the Planck function $ B_\nu$ and as a harmonic mean it emphasizes the contributions from the least opaque (smallest opacity) windows. It therefore reproduces the correct flux in the optically thick layers, exactly where brown dwarf interiors inject their heat. Replacing the constant, i.e., gray opacity of the double-gray method and using a single mean opacity rather than tens of thousands of wavelength bins used for multi-wavelength methods, we both preserve the essential $T\text{--}p$ dependence and achieve this at a reasonable computational cost. It should be noted that while the Rosseland mean delivers an accurate depiction of the diffusive and optically thick limit, in the optically thin parts of the upper atmosphere, the Planck mean should be preferred \citep{marley_2015}. In our single-band gray scheme, we do not switch means; following prior work by \citet{tan_2021b}, \citet{komacek_2022_clouds} and \citet{lee_2023_BD}, we impose a gray lower limit of $\kappa_{\mathrm{R, min}} = 10^{-3}~ \rm m^{2}~kg^{-1}$.

We expect the gas-phase thermal scattering to be negligible \citep{lee_2023_BD}, but cloud particles can contribute appreciably. For this reason, we implement a further modification (detailed in Section \ref{sec:thermal_feedback_and_scattering_of_clouds}) to account for the scattering induced by the presence of clouds in our models and only activate it when clouds are present.

\subsubsection{Thermal forcing of the atmosphere}
\label{sec:thermal_forcing_of_the_atmosphere}

In our curated sample, with no instellation present, the dynamical response of the atmosphere is triggered through variations in the internal thermal flux coupling to the brown dwarfs' rotation. Conceptually, the physical scenario we aim to model corresponds to convective plumes that rise and sink at the RCB. Several authors \citep[e.g., ][]{zhang_showman_2014} have modelled this mechanism and \citet{showman_tan_zhang_2019} have outlined their approach that works through stochastic thermal perturbations introduced at the RCB, which we follow in our current work.

Briefly summarized, thermal perturbations are characterized through functional forms $S\left(\lambda, \phi, p, t \right)$ that are separated into the vertical $S_{\text{v}}\left(p\right)$ and horizontal $S_{\text{h}}\left( \phi, p, t \right)$ terms, where $\lambda$ is longitude, $\phi$ is latitude, $p$ is pressure, and $t$ is time. The vertical structure is designed to decay over two scale heights in order to keep the perturbations localized near the RCB. In our study, this level is set to a constant $p_{\mathrm{rcb}} = 10~\rm bar$, matching other recent modelling efforts \citep{lee_2024_BD_clouds}.

Physically, the above-mentioned buoyant plumes are present at a much smaller scale than our grid scale \citep{freytag_2010} and are introduced at the smallest resolved scale in our GCM setup. This implementation implicitly assumes that the smaller-scale plumes reorganize themselves into a continuous larger-scale structure corresponding to our grid scale \citep{showman_tan_zhang_2019}. Finally, using their physically motivated description, we estimate the amplitude of the thermal perturbations as,
\begin{equation}
    T_{\mathrm{amp}} \approx \frac{\Delta z}{H}\frac{T_{\mathrm{eff}}}{\tau_{\mathrm{rad}} \sqrt{n_{f}}},
\end{equation}
where $\Delta z$ is the maximum amplitude of the displacement, $H$ the pressure scale height, $T_{\mathrm{eff}}$ the effective temperature, $\tau_{\mathrm{rad}}$ the radiative timescale and $n_{f}$ the forcing wavenumber. This completes the physical picture as we have chosen the relevant length scale and the amplitude of the perturbations we would like to introduce. The horizontal structure, then, is represented as a Markov process, i.e., a sequence of stochastic events that depend on the previous state,
\begin{equation}
\label{eq:thermal_forcing_markov}
    S\left(\lambda, \phi, p, t +\delta t \right) = r S\left(\lambda, \phi, p, t \right) + \sqrt{1 - r^{2}} F\left(\lambda, \phi, t \right),
\end{equation}
which has its roots in turbulence studies \citep{scott_forced-dissipative_2007, showman_tan_zhang_2019}. This parametrization provides the simplest stochastic representation, in the absence of a detailed physical model, consistent with the idea that convective events at the RCB persist for a short duration before being replaced by new plumes. 

To give a conceptual explanation, Eq. \ref{eq:thermal_forcing_markov} updates the horizontal forcing pattern by combining two contributions. The first term, $r S(\lambda,\phi,p,t)$ with $r = 1 - \frac{\delta t}{t_{\mathrm{storm}}}$, retains part of the previous pattern, such that $t_{\rm storm}$ sets the timescale over which the forcing loses memory of its past. Since $\delta t \ll t_{\rm storm}$, we have $0 < r < 1$, so the pattern evolves gradually from one step to the next. The second term introduces new variability through $F(\lambda,\phi,t)$ constructed as a sum of spherical harmonics with amplitude set by $T_{\rm amp}$ and a characteristic horizontal length scale set by the forcing wavenumber $n_{f}$. The prefactor $\sqrt{1-r^2}$ scales the strength of this stochastic contribution so that the forcing neither decays nor grows without bound over many updates. Given a sufficient number of iterations, this yields a continually refreshed pattern with the prescribed horizontal scale, intended to represent the aggregate effect of unresolved convective plumes on the resolved flow. Our chosen parameters match the \citet{lee_2024_BD_clouds} study and are summarized in Table \ref{tab:cloud_numerics}.

\subsubsection{Mixing length theory}
\label{sec:mixing_length_theory}

To model the convective interior of brown dwarfs, we introduce a mixing-length scheme after \citet{marley_2015}, \citet{joyce_tayar_2023} and \citet{lee_2024_BD_clouds}. Mixing length theory (MLT) \citep{prandtl_1925, boehm_vitense_1958} provides us with a parametric approach for representing the convective mixing that happens on a sub-grid scale.

Physically, the modelled scenario can be described as a parcel of fluid, rising buoyantly in the vertical direction due to its excess temperature, in regions where the atmosphere is unstable to convection. The stability of a region of the atmosphere is assessed through the Schwarzschild \citep{schwarzschild_1906} criterion,
\begin{equation}
    \Gamma \geq \Gamma_{\mathrm{ad}},
\end{equation}
where $\Gamma = -\frac{dT}{dz}$ is called the lapse rate and $\Gamma_{\mathrm{ad}} = \frac{g}{c_{p}}$ is the adiabatic lapse rate, i.e., the rate that the temperature would decrease with altitude without exchanging heat with its surroundings. The core assumption of mixing-length theory is that a displaced parcel exchanges heat with its surroundings after travelling a characteristic distance $L$, called the mixing length. The mixing length is approximated as a multiple of the scale height $H$ as $L = \alpha H$, where $\alpha$ is an order-unity constant. We adopt $\alpha = 1$, following \citet{lee_2024_BD_clouds}, while noting that the methodology of choosing of this parameter is still actively debated \citep{smith_estimation_1998, joyce_tayar_2023}. For a detailed discussion of how adjusting this free parameter influences the results, we refer the reader to \citet{robinson_marley_2014}.

MLT is inherently a one-dimensional approximation, where the mixing is assumed to occur only in the vertical direction. The mixing between the rising fluid parcel and its new surroundings results in a convective flux, 
\begin{equation}
\label{eq:convective_flux}
    F_{\text{conv}} = \frac{1}{2} \rho c_{p} w L \left(\Gamma - \Gamma_{\text{ad}} \right),
\end{equation}
where $\rho$ is bulk density of the atmosphere, $c_{p}$ the specific heat capacity under constant pressure and $w$ the vertical mixing velocity \citep{marley_2015},
\begin{equation}
\label{eq:w_MLT}
    w = L \left[ \frac{g}{T} \left(\Gamma - \Gamma_{\text{ad}} \right) \right]^{1/2}.
\end{equation}

Eq. \ref{eq:convective_flux} is sometimes written in terms of the convective eddy diffusivity, approximated as, $K_{zz} = wL$ \citep{marley_2015}. The convective heat flux $F_{\text{conv}}$ is used to update the local temperature tendencies in every computational cell according to,
\begin{equation}
    \left(\frac{\partial T}{\partial t} \right)_{\text{conv}} = \frac{-1}{\rho c_{P}}\frac{\partial F_{\text{conv}}}{\partial z},
\end{equation}
using a sub-timestepping approach. We adopt an MLT timestep of $\Delta t_{\mathrm{MLT}} = 0.5~\rm s$. Once the cumulative MLT updates reach the main dynamical timestep, the resulting temperature correction is passed to the dynamical solver as an update to the heating rate $Q_{\mathrm{heat}}$.

As described above, this approach yields convective heat fluxes and, in turn, a local and time-varying thermal eddy diffusivity coefficient
\begin{equation}
    \label{eq:kzz}
    K_{zz} = L^2 \left[ \frac{g}{T} \left(\Gamma - \Gamma_{\text{ad}} \right) \right]^{1/2},
\end{equation}
evaluated in each layer that meets the instability criterion. This usage of $K_{zz}$ is local and time-dependent and differs from the globally-averaged eddy diffusivity coefficients sometimes adopted in theoretical or modeling studies \citep[e.g.,][]{steinrueck_2021, tan_2022}. Our approach yields a self-consistent mixing profile that can distinguish vertically separated convective cells. While not as accurate as convective-resolving models \citep[such as][]{lefevre_2022_convection_BD}, the scheme captures the key features of interior mixing that we can couple to our cloud tracer scheme detailed in Section \ref{sec:modelling_of_clouds}. At the same time, the MLT routine can only capture mixing where the convective heat flux is nonzero ($F_{\mathrm{conv}}~\textgreater~0$), while the vertical velocity above a convective zone decays rather than dropping to zero. This "overshoot" effect has been noted by multiple studies relevant for brown dwarf atmospheres \citep{freytag_2010,lefevre_2022_convection_BD, lee_2024_BD_clouds}. We therefore extrapolate the velocity above the top (i.e., to pressures lower than $p_{\mathrm{top}}$) of each convective zone with the prescription of \citet{woitke_2020},
\begin{equation}
    \ln w_{\mathrm{ov}} = \ln w_{\mathrm{top}} - \beta \cdot \max \left(0, \ln p_{\mathrm{top}} - \ln p \right),
\end{equation}
where $\beta$ is a free parameter that relates to \citet{freytag_2010} and adopted as $\beta = 2.2$, following the work of \citet{lee_2024_BD_clouds} in our study, and $w_{\mathrm{top}}$ represents the vertical velocity at the top of a given convective cell, as determined using Equation \ref{eq:w_MLT}. This overshoot prescription feeds directly into convective $K_{zz}$ structure via $K_{zz, \mathrm{tot}} = K_{zz} + K_{zz, \mathrm{ov}}$ where $K_{zz, \mathrm{ov}} = w_{\mathrm{ov}} L $, giving a smooth decline of $K_{zz}$ in the stably stratified region above each convective boundary. The overshoot term $K_{zz, \mathrm{ov}}$ is applied solely as a correction for tracer mixing (detailed in Section \ref{sec:modelling_of_clouds}) and does not affect the MLT temperature tendencies (through $Q_{\mathrm{heat}}$), since $F_{\mathrm{conv}} = 0$ in a stable region.
 
\subsubsection{Modelling of cloud tracers}
\label{sec:modelling_of_clouds}
To probe the effect of clouds on the thermal structure and the ensuing feedback, we implement a simple tracer-based equilibrium cloud scheme. Our implementation closely follows the works of \citet{tan_atmospheric_2019}, \citet{tan_2021a}, \citet{komacek_2022_clouds} and \citet{lee_2024_BD_clouds}. The model evolves the following equations:

\begin{align}
    \frac{Dq_{v}}{dt} &=
        (1-s)\frac{\min(q_{s}-q_{v},\,q_{c})}{\tau_{c}}
      - s\frac{q_{v}-q_{s}}{\tau_{c}}                                      \nonumber\\
    &\hspace{2em}
      -\frac{q_{v}-q_{\text{deep}}}{\tau_{\text{deep}}}
      +\frac{1}{\rho}\frac{\partial}{\partial z}
         \left( K_{zz}\,\rho\,\frac{\partial q_{v}}{\partial z}\right),
    \label{eq:vapour_tracer}\\[0.5em]
    \frac{Dq_{c}}{dt} &=
        s\frac{q_{v}-q_{s}}{\tau_{c}}
      -(1-s)\frac{\min(q_{s}-q_{v},\,q_{c})}{\tau_{c}}                     \nonumber\\
    &\hspace{2em}
      -\frac{1}{\rho}\frac{\partial}{\partial z}
         \bigl(\rho\langle q_{c}V_{s}\rangle\bigr)
      +\frac{1}{\rho}\frac{\partial}{\partial z}
         \left( K_{zz}\,\rho\,\frac{\partial q_{c}}{\partial z}\right).
    \label{eq:condensate_tracer}
\end{align}
where $q_{v}$ is the vapor mass-mixing ratio (MMR), $q_{c}$ the condensate MMR, $q_{s}$ the equilibrium saturation ratio, $\tau_{c}$ the condensation timescale, $q_{\text{deep}}$ the deep reservoir MMR, $\tau_{\text{deep}}$ the deep mixing timescale and $s$ the saturation coefficient.
Conceptually, this is an equilibrium condensation model with a prescribed relaxation timescale applied to the tracers, while advection of tracers follows the bulk flow of the atmosphere passively. For the relaxation timescale, we follow \citet{tan_atmospheric_2019} , who prescribe $\tau_{\mathrm{deep}}=10^{3}~\rm s$ and demonstrate that varying $\tau_{\mathrm{deep}}$ by orders of magnitude has little effect. The condensation timescale $\tau_{c}$ follows \citet{lee_2024_BD_clouds} (see Table~\ref{tab:cloud_numerics}), who adopt $\tau_{c} = 120~\mathrm{s}$. This is also consistent with \citet{lefevre_2022_convection_BD}, who use a smaller value ($\tau_{c}= 10~\mathrm{s}$) and find that varying $\tau_{c}$ from $1~ \rm s$ to $100~\mathrm{s}$ produces only small changes in the cloud layer depth. For the stability criterion, we adopt the convention of \citet{lee_2024_BD_clouds}, defining it in terms of the supersaturation ratio S, given by
\begin{equation}
    S = \frac{\chi_v}{\chi_s},
\end{equation}
where $\chi_s$, the equilibrium VMR, is computed as $\chi_s = \min\left(p_{\mathrm{vap}}(T)/p, 1 \right)$, with $p_{\mathrm{vap}}(T)$ the saturation vapor pressure as a function of temperature. The saturation coefficient $s$ is then defined as
\begin{equation}
    s =
        \begin{cases}
        1, & \text{if } S > 1 \\
        0, & \text{if } S < 1 \\
        \frac{dq_v}{dt} = \frac{dq_c}{dt} = 0, & \text{if } S = 1
        \end{cases}
\end{equation}

This ensures that phase changes only occur when the system is either supersaturated (condensation) or subsaturated (evaporation), and that no mass exchange occurs in equilibrium (S = 1). Inside the GCM, this set of equations is represented through a standard continuity equation with a source term in the following form,
\begin{equation}
      \frac{\partial \rho q}{\partial t} + \nabla \cdot \left( \rho \mathbf{v} q \right) = \mathcal{S}, \label{eq:continuity_eq_w_source}
\end{equation}
where $\rho$ is the bulk density of the background atmosphere, $q$ represents the MMR of either the condensable vapor or the condensate of a given cloud species, $\mathbf{v}$ the wind field, and $\mathcal{S}$ the source and sink terms for each equation. 

The left-hand side of equation \ref{eq:continuity_eq_w_source} has the same structural form as the entropy equation evolved by the THOR dynamical core and is solved using the same finite-volume scheme introduced in \citet{mendonca_thor_2016} and \citet{deitrick_thor2.0_2020}. A more detailed derivation of how equations \ref{eq:vapour_tracer} and \ref{eq:condensate_tracer} relate to equation \ref{eq:continuity_eq_w_source} is given in Appendix \ref{sec:appendix_adaptation_of_the_tracer_equations}. The source terms, represented by the term $\mathcal{S}$ on the right-hand side, are integrated separately with a sub-timestepping approach. Specifically, the treatment of source terms is split into three components. The first part, the mass exchange between vapor and condensate phases, along with vapor replenishment from the deep reservoir is integrated using a third-order adaptive Bogacki–Shampine Runge–Kutta scheme \citep{bogacki_32_1989}. Similarly, the vertical mixing terms calculated with the $K_{zz}$ values derived from the MLT routine are integrated using an adaptive Bogacki-Shampine routine. We use the third-order Bogacki–Shampine method for both cases because lower-order integrators were not stable at reasonable time step sizes. Additionally, its embedded second-order solution gives a cheap error estimate for adaptive time stepping, which avoids a conservative fixed time step that would bottleneck performance and saves the time otherwise spent searching for suitable step values. The treatment of the gravitational settling of condensate particles requires a more detailed approach and is discussed in the following subsection.

The physical scenario we model assumes that there is a deep $\left( p > 100~ \rm bar \right)$ vapor reservoir that has a solar elemental composition. For each condensate, we set $q_{\text{deep}}$ (see Table \ref{tab:cloud_numerics}) using the solar number density ratios $\left(n\left(\mathrm{El}\right)/n\left(\mathrm{H}\right) \right)$ of \citet{asplund_2021_solar_abundances}, taking the value for the molecule’s least abundant constituent, which serves as the bottleneck in forming the condensate. Our assumption is guided by the fact that most Y dwarfs are observed in the Solar neighborhood, and have metallicities close to solar \citet{legget_2017}. We adopt a solar-abundance deep reservoir when no stronger observational or first-principles constraints are available, following the conventions of \citet{gao_benneke_2018_clouds}, \citet{komacek_2022_clouds} and \citet{lee_2024_BD_clouds}. At the beginning of our simulations, we set the condensate MMRs to a very small value, $q_{c} = 10^{-30}$, while the corresponding vapor MMRs follow their saturation curves but never exceed their respective $q_{\mathrm{deep}}$. In other words, we assume the deep reservoir has already come into equilibrium with the overlying atmosphere before cloud formation begins.

\subsubsection{Gravitational settling of clouds}
\label{sec:gravitational_settling_of_clouds}

Equations \ref{eq:vapour_tracer} and \ref{eq:condensate_tracer} provide a simplified framework that permits mass exchange between vapor and condensate. In equilibrium, the system maintains a delicate balance between the upward mixing from the deep vapor reservoir and the gravitational settling of cloud condensates. The resulting large-scale cloud structure then depends on the efficiency of vertical mixing, both advective and convective, throughout the atmosphere.

Because we do not use an explicit microphysical model, assumptions about the cloud particle size distribution must be made to estimate settling velocities. In the absence of a first-principles theory, we follow convention and adopt a prescribed log-normal size distribution \citep{ackerman_marley_2001}
\begin{equation}
    \frac{dn}{dr} = \frac{N_{0}}{r \sqrt{2 \pi} \ln \sigma_{g}} \exp \left( - \frac{\ln \left( r/ r_{\mathrm{med}} \right)^{2}}{2 \ln \sigma_{g}^{2}}\right),
\end{equation}
defined by a median particle radius $r_{\mathrm{med}}$ and a fixed geometric standard deviation $\sigma_{g}$, to parameterize the bulk settling behaviour of condensates. This can be integrated to yield the total number density
\begin{equation}
\label{eq:number_density}
    N_{0} = \frac{3 \epsilon q_{c} \rho}{4 \pi \rho_{c} r_{\mathrm{med}}^{3}} \exp \left(-\frac{9}{2} \ln^{2} \sigma \right),
\end{equation}
where $\rho_{c}$ is the density of the condensate particle and $\epsilon =\mu_{c} / \mu_{\mathrm{gas}}$ is the molecular weight ratio between the condensate and the background gas. In this framework, the effective cloud particle radius can be expressed as \citep{lee_2024_BD_clouds},
\begin{equation}
    r_{c} = r_{\mathrm{med}} \exp \left( \frac{7}{2} \ln^2 \sigma \right).
\end{equation}
This corresponds to the \citet{ackerman_marley_2001} expression (Eq. 13 in their work) with $f_{\mathrm{rain}} = 1$ and $\alpha = 1$. Having defined the size distribution of cloud condensates, we estimate the settling velocity through \citep{ohno_okuzumi_2018, lee_ohno_2025}
\begin{equation}
\label{eq:settling_velocity}
    V_{s} = \frac{2 \beta g r_{c}^{2}}{9 \eta} \left[ 1+ \left(\frac{0.45 g r_{c}^{3} \rho_{c} \rho}{54 \eta}\right)^{2/5} \right]^{-5/4},
\end{equation}
where $g$ is surface gravity, $\eta$ the dynamic viscosity, $\rho_{c}$ the density of the condensate and $\rho$ the bulk density of the background atmosphere. This expression is an extension of the standard expression for terminal velocity \citep{rossow_1978}, derived assuming a Stokes flow, for a particle falling in a viscous flow. The empirical factor $\beta$, called the Cunningham slip factor \citep{cunningham_1910, davies_1945},
\begin{equation}
    \beta = 1 + Kn~ \left(1.257  + 0.4 ~ e^{-1.1/ Kn} \right),
\end{equation}
accounts for deviations from continuum behaviour in the flow and is a function \citep{li_wang_2003} of the Knudsen number ,
\begin{equation}
    Kn = \frac{\lambda}{r_{c}},
\end{equation}
where $\lambda$ is the mean free path of the fluid, calculated as \citep{jacobson_2005}
\begin{equation}
    \lambda = \frac{2 \eta}{\rho} \sqrt{\frac{\pi \mu}{8 R T}},
\end{equation}
and $\eta$ is the dynamical viscosity according to \citet{rosner_2000}
\begin{equation}
    \eta = \frac{5}{16} \frac{\sqrt{\pi \mu m_{u} k_{\mathrm{B}} T}}{\pi d_{\mathrm{H_{2}}}} \frac{\left(\frac{k_{\mathrm{B}} T}{\epsilon_{\mathrm{LJ, H_{2}}}}\right)^{0.16}}{1.22},
\end{equation}
where $m_{u}$ is the atomic mass unit, $d_{\mathrm{H_{2}}}$ the molecular diameter of $\mathrm{H_{2}}$ and $\epsilon_{\mathrm{LJ,H_{2}}}$ the Lennard-Jones potential of $\mathrm{H_{2}}$. \citet{lee_ohno_2025} adopt the same expression, while they consider contributions $\mathrm{H_{2}}, \mathrm{H}$ and $\mathrm{He}$. In this study, we limit ourselves to $\mathrm{H_{2}}$ for the sake of simplicity. Numerically, we calculate the vertical transport using a second-order explicit MacCormack \citep{maccormack_effect_2001} integrator with a Koren \citep{koren_1993} flux limiter. 

\subsubsection{Thermal feedback and scattering due to clouds}
\label{sec:thermal_feedback_and_scattering_of_clouds}

So far, we have described how cloud tracers are distributed throughout the atmosphere, generally following the bulk motion of the background gas, while accounting for mass exchange due to evaporation and condensation, mixing from convective motions, and gravitational settling of condensates. As mentioned in Section \ref{sec:introduction}, previous studies \citep{morley_2012} indicate that for Y dwarf atmospheres, clouds are expected to drive dynamical activity. This necessitates that we allow the cloud distribution to thermally feed back on the background atmosphere. 

Physically, to capture the radiative impact of clouds, we must specify the optical depth and scattering properties of each cloud species. The optical depth is computed from specifying the extinction opacity of clouds $\kappa_{\mathrm{ext}}~[\mathrm{m^{2}~kg^{-1}}]$. The scattering properties of the cloud particles are described by the single scattering albedo, $\omega~[\text{--}]$, which quantifies the fraction of incident radiation that is scattered,  rather than absorbed by cloud particles, and the asymmetry factor, $g~[\text{--}]$, which characterizes the directional bias of scattering, i.e., the ratio of forward to backward scattering. These quantities, however, are wavelength-dependent and require knowledge of the size distributions of cloud particles. To integrate them into our radiative transfer scheme, which uses a single thermal band for brown dwarf emission, we follow the procedure described in \citet{lee_2024_BD_clouds}, adopting the size distributions and parameters detailed in Section \ref{sec:gravitational_settling_of_clouds}. We use the LX-MIE \citep{kitzmann_2018_lxmie} code to pre-calculate the wavelength-dependent cloud properties and, assuming a lognormal size distribution, integrate over size to obtain the normalized attenuation coefficient $\alpha~[\mathrm{m^{-2}}]$, the single-scattering albedo $\omega~[\text{--}]$ and the asymmetry factor $g~[\text{--}]$. We then form Rosseland-mean values by weighting with $\partial B_{\nu}/\partial T$ on a temperature grid and tabulate the resulting normalized attenuation coefficient $\alpha_{\mathrm{R}}(T)$, single-scattering albedo $\omega_{\mathrm{R}}(T)$, and asymmetry factor $g_{\mathrm{R}}(T)$, which we interpolate to local background gas temperature at runtime. This allows us to calculate, on the fly, the Rosseland mean cloud opacity of a single cloud species as 
\begin{equation}
    \kappa_{\mathrm{R, cloud}} = \frac{\alpha_{\mathrm{R}} N_{0}}{\rho},
\end{equation}
where the number density, $N_{0}~[m^{-3}]$ is given by Eq. \ref{eq:number_density}. We approximate the contribution of our cloud species' thermal contribution by adding it to the background gas opacity, interpolated from the tabulated \citet{freedman_2014} opacities (as described in \ref{subsection:2s_nongray_IR_RT}),
\begin{equation}
    \kappa_{\mathrm{R, tot}} = \kappa_{\mathrm{R, gas}} + \sum_{i}^{n_{\mathrm{cloud}}}\kappa_{\mathrm{R, cloud}, i}. 
\end{equation}

It should be noted that, although used in several other studies \citep{tan_2021b, komacek_2022_clouds, lee_2023_BD}, this is an approximation of total opacity. We apply a further correction to account for scattering introduced by the cloud particles present at a given layer of the atmosphere, through the absorption approximation of \citet{lee_2024}. The idea behind this approximation is to account for scattering through a modification of the transmission function during our radiative transfer calculations. The total single scattering albedo $\omega_{\mathrm{R, tot}}$ and the total asymmetry factor $g_{\mathrm{R, tot}}$ are calculated as,
\begin{align}
    \omega_{\mathrm{R, tot}} & = \frac{\omega_{\mathrm{R, gas}} \kappa_{\mathrm{R, gas}} + \sum_{i}^{n_{\mathrm{cloud}}} \omega_{\mathrm{R, cloud}, i} \kappa_{\mathrm{R, cloud}, i}}{\kappa_{\mathrm{R, tot}}}, \\
    g_{\mathrm{R, tot}} & = \frac{g_{\mathrm{R, gas}} \omega_{\mathrm{R, gas}} \kappa_{\mathrm{R, gas}} + \sum_{i}^{n_{\mathrm{cloud}}} g_{\mathrm{R, cloud}, i} \omega_{\mathrm{R, cloud}, i} \kappa_{\mathrm{R, cloud}, i}}{\omega_{\mathrm{R, tot}} \kappa_{\mathrm{R, tot}}}.
\end{align}

For each layer, we first apply the $\delta \text{-M}+$ scaling of \citet{lin_new_2018}, to remove the strong forward-scattering peak and keep the two-stream calculation accurate. The resulting modified single-scattering albedo and optical depth are denoted as $\omega_{\mathrm{R, tot}}^{\star}$ and $\tau^{\star}$. The modified co-albedo is then
\begin{equation}
  \epsilon^{\ast} =  \sqrt{\left(1-\omega_{\mathrm{R, tot}}^{\star}\right) \left(1-g_{\mathrm{R, tot}}\omega_{\mathrm{R, tot}}^{\star}\right)},
  \label{eq:co_albedo}
\end{equation}
which multiplies the layer optical depth to give the absorption optical depth $\tau_{\mathrm{a}}=\epsilon^{\ast}\tau^{\star}$. Our radiative transfer solver (see Section \ref{subsection:2s_nongray_IR_RT}) then evaluates the transmission function,
\begin{equation}
  \mathcal{T}(\mu) = \exp \left[-\tau_{\mathrm{a}}/\mu \right],
  \label{eq:transmission_function}
\end{equation}
for every Gaussian angle $\mu$. \citet{lee_2024} conclude that this method provides a good approximation of the scattering behaviour while coming at negligible computational cost.

With these modifications in place, initially, we turn cloud tracers on passively, allowing them to evolve without contributing to the thermal budget. After 250 Earth days, sufficient time for the dynamics to redistribute the tracers and for condensates to settle into statistically steady cloud layers, we switch on their thermal feedback. Starting this feedback later avoids spurious heating at the lower boundary. The reasoning behind this is that the initial state is saturated yet condensation has not yet occurred, tracer concentrations are initially greatest near the surface, and immediate coupling would artificially amplify their effect there.

\section{Results}
\label{sec:results}
\subsection{Atmospheric structure of the Y dwarf sample}

\begin{figure*}[t!]
    \centering
    
    \begin{subfigure}[b]{\textwidth}
        \centering
        \includegraphics[width=0.85\textwidth]{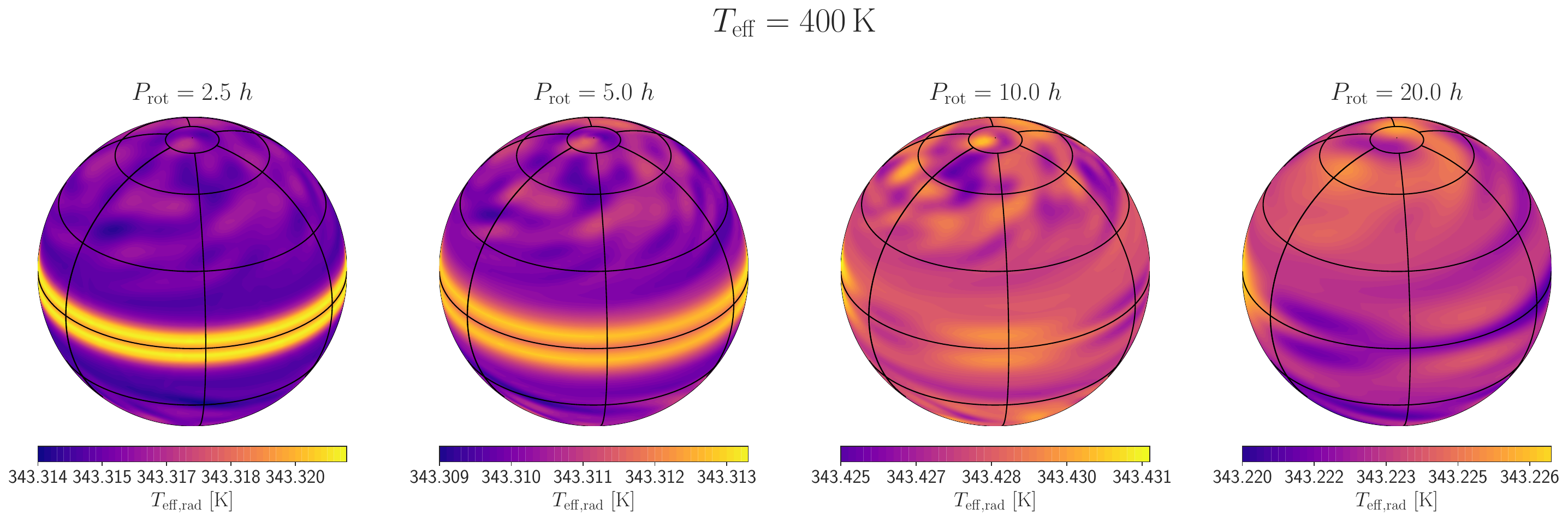}
        \caption{}
        \label{subfig:olr_400K}
    \end{subfigure}
    \vspace{0.4em}

    \begin{subfigure}[b]{\textwidth}
        \centering
        \includegraphics[width=0.85\textwidth]{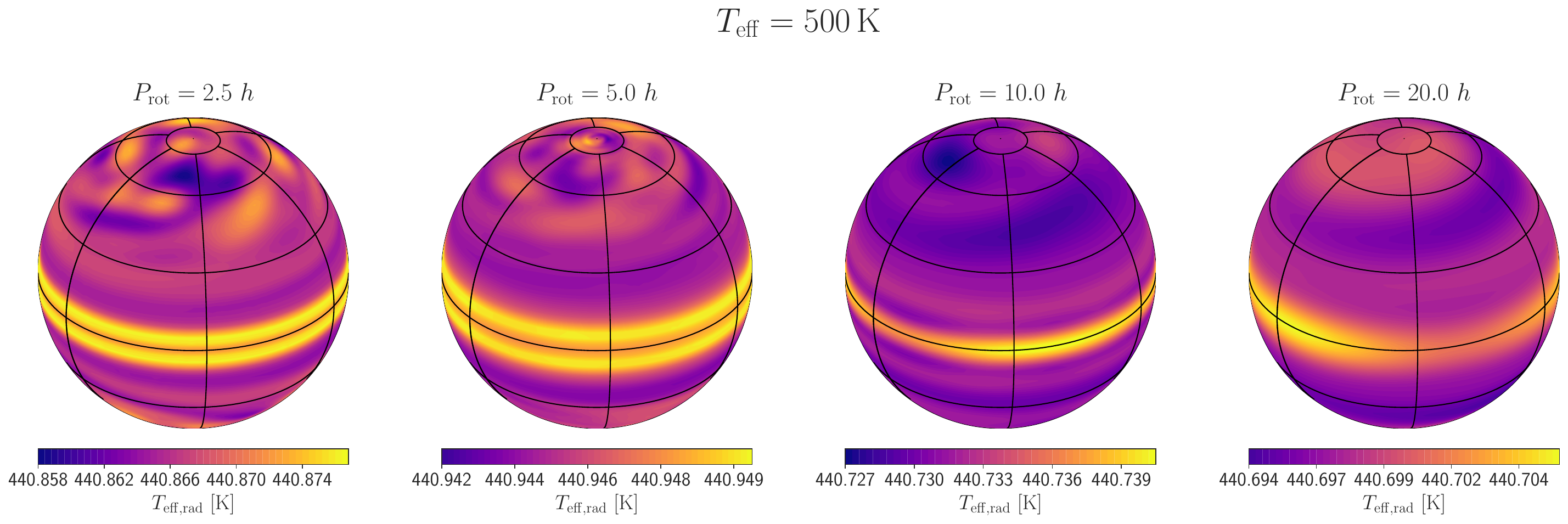}
        \caption{}
        \label{subfig:olr_500K}
    \end{subfigure}
    \vspace{0.4em}

    \begin{subfigure}[b]{\textwidth}
        \centering
        \includegraphics[width=0.85\textwidth]{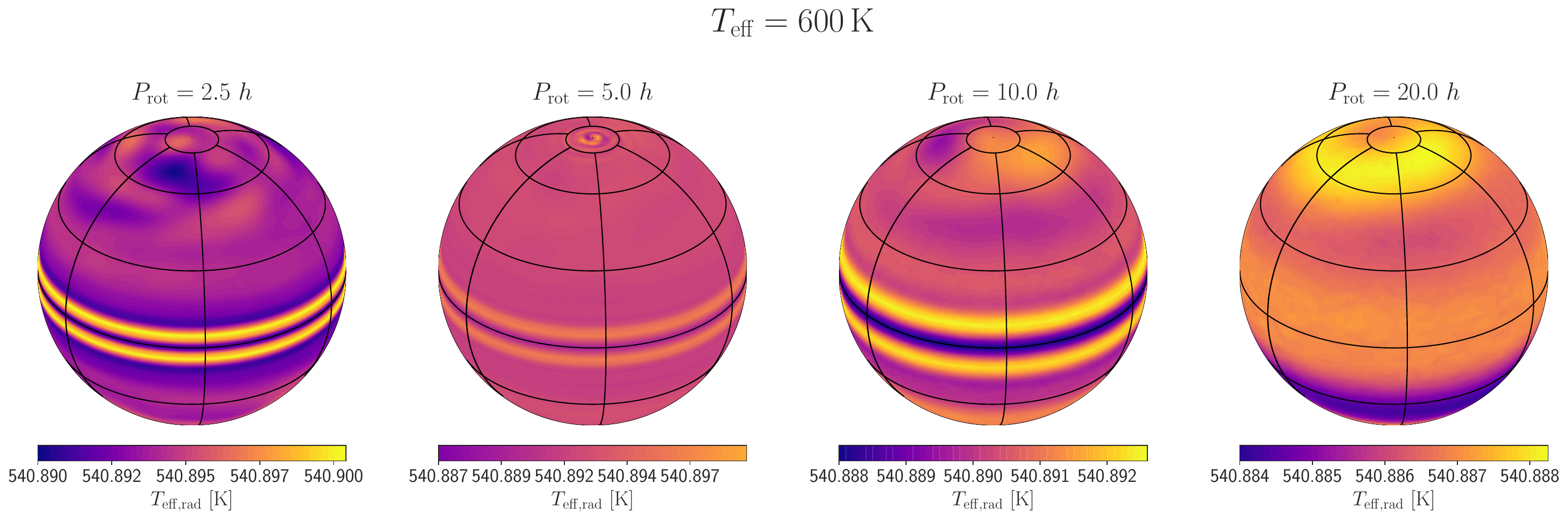}
        \caption{}
        \label{subfig:olr_600K}
    \end{subfigure}

    \caption{Effective radiating temperature maps of individual runs at $t_{\mathrm{run}} = 1000~d$. The shown results are time-averaged over the last 10 days of the given run time. Run cases vary in effective temperature from $T_{\mathrm{eff}} = 400~K$ to $T_{\mathrm{eff}} = 600~\rm K$ from top to bottom and correspond to a rotation period of $P_{\mathrm{rot}} = 2.5 ~\text{--}~ 20~ \rm h$ from left to right.}
    \label{fig:olr_grid}
\end{figure*}

Having constructed a broad suite of Y dwarf simulation cases, we now examine the steady-state behavior of each run after reaching statistical equilibrium. Figure \ref{fig:olr_grid} presents the full sequence of simulations, with each panel showing the effective radiating temperature $T_{\text{eff, rad}}$ for a given combination of effective temperature $T_{\mathrm{eff}}$ and rotation period $P_{\mathrm{rot}}$. The effective radiating temperature is obtained from the outgoing longwave radiation (OLR) flux via $F_{\mathrm{OLR}} = \sigma_{\mathrm{SB}}T_{\text{eff, rad}}^4$, providing a temperature-equivalent measure of the gray outgoing flux emerging from the photosphere. Across cases of the same effective temperature, it varies only slightly, typically by less than $1\%$. While equatorial features are prominent in all simulations, their detailed morphology differs. For $T_{\mathrm{eff}} = 400~\mathrm{K}$, faster rotators exhibit slightly enhanced equatorial flux, with patchy structures appearing at higher latitudes in all cases. For the slower rotating $P_{\mathrm{rot}} = 10~\mathrm{h},~20~\mathrm{h}$ cases at $T_{\mathrm{eff}} = 400~\mathrm{K},500~\mathrm{K}$, an oscillatory equatorial pattern emerges, with small-scale, high-latitude variability. The $T_{\mathrm{eff}} = 600~\mathrm{K}$ cases appear more uniform from the equator to mid-latitudes, while the polar regions show increased patchiness. As expected, higher resolution and faster rotation generally lead to finer-scale atmospheric features, although the most significant jump is from the $P_{\mathrm{rot}} = 20~\rm h$ to the faster rotators, which display similar characteristic length scales in the structures that end up forming. Although Figure \ref{fig:olr_grid} may suggest that many $T_{\mathrm{eff}}=500~\mathrm{K}, 600~\mathrm{K}$ cases are uniform away from the equator, this impression arises because equatorial variability is much stronger than at higher latitudes, as the $T_{\mathrm{eff}}= 500 ~\mathrm{K}~;~P_{\mathrm{rot}}=2.5~\mathrm{h}$ case indicates, high-latitude vortices are present in all runs but are too weak to stand out against the equatorial signal. Both the $T_{\mathrm{eff}} = 400~\mathrm{K}$ and $500~\mathrm{K}$ cases develop oscillatory equatorial patterns, with the $500~\mathrm{K}$ runs showing the largest contrast and strongest winds, whereas at $T_{\mathrm{eff}} = 600~\mathrm{K}$ the equatorial variability weakens and the flow is dominated by broad zonal bands, with variability shifted toward the polar regions.

Overall, the effective radiating temperature maps in Figure \ref{fig:olr_grid} are close to horizontally uniform, with only very small peak-to-peak contrasts, while still exhibiting coherent banded and tilted structures indicative of three-dimensional circulation. The small horizontal variations in these maps are consistent with a weak dynamical regime, which we discuss further in Section \ref{subsection:flow_regime}.

Figure \ref{fig:Kzz_composite} illustrates the evolution of the temperature–pressure ($T\text{–}p$) profiles across the entire set of simulations. As discussed in Section \ref{sec:run_cases}, the final $T\text{–}p$ profiles remain remarkably close to the initial conditions shown in Figure \ref{fig:eddington_profs}. A weak inversion forms early in the simulation (not shown), but the atmosphere settles into a structure featuring an adiabatic interior and a nearly isothermal upper atmosphere extending to pressures as low as $p \approx 10^{-4}~\mathrm{bar}$. The profiles shown are global averages, but individual columns reveal negligible horizontal temperature variation, consistent with the absence of any imposed longitudinal or latitudinal asymmetry in thermal forcing.

\subsection{Vertical mixing}

\begin{figure*}
    \centering
    \includegraphics[width=0.9\textwidth]{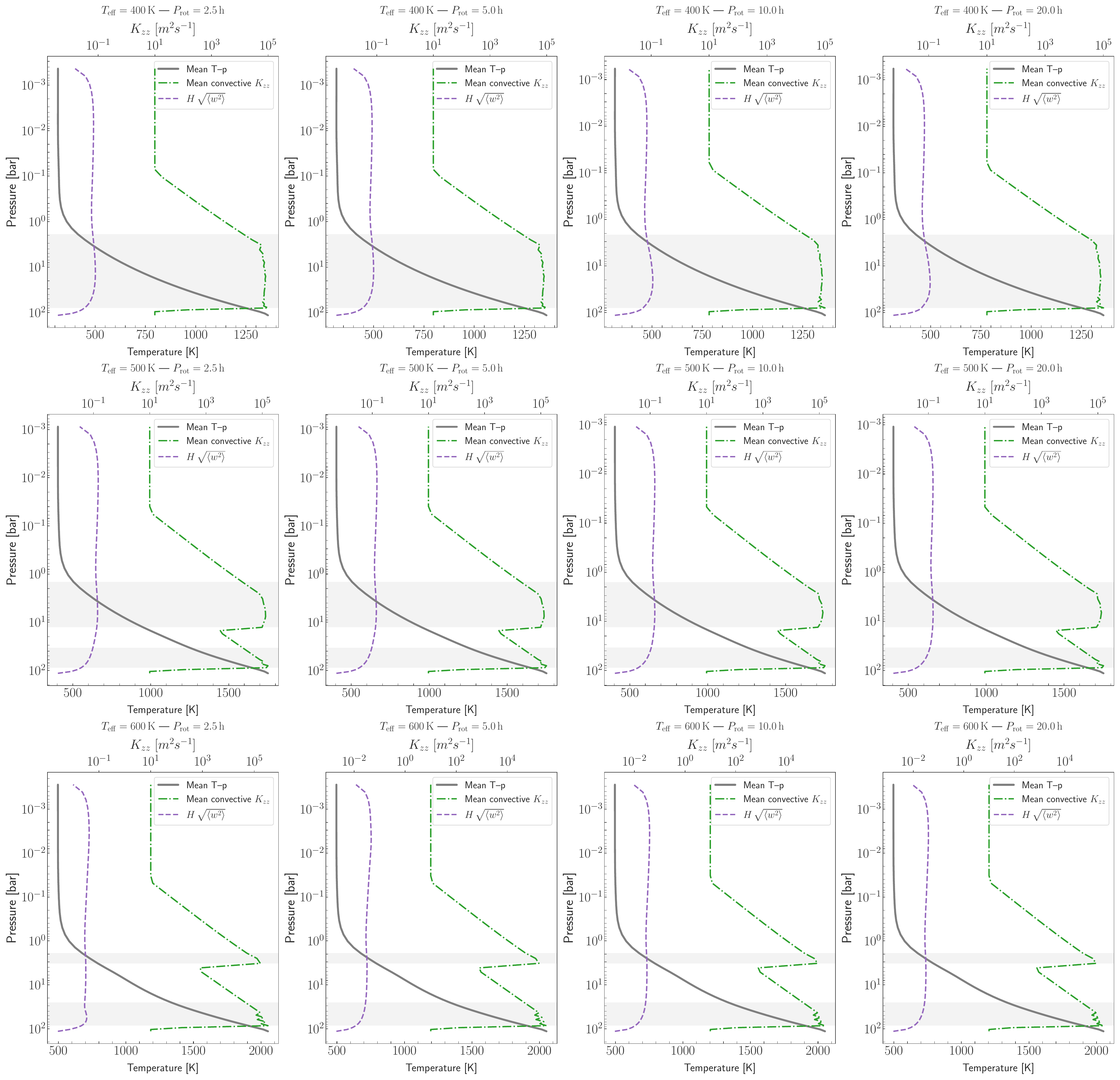}
    \caption{Vertical mixing profiles of all simulations. The $T\text{--}p$ profile is overplotted with the convective mixing $K_{zz}$ and the advective mixing contribution $H \left(z \right) \sqrt{\langle w^{2} \left( z \right ) \rangle}_{h}$. The pressure extents of convective regions are shown as gray horizontal bands in each panel.}
    \label{fig:Kzz_composite}
\end{figure*}

We expect vertical mixing to be one of the main drivers of dynamics in Y dwarf atmospheres. In Figure \ref{fig:Kzz_composite}, we overplot the $T \text{--} p$ structure with the globally-averaged eddy diffusivities $K_{zz}$, calculated according to Eq. \ref{eq:kzz} by the MLT routine, giving a measure of the convective mixing that is going on in the atmosphere, and the "advective" mixing that is calculated from the GCM's own vertical velocity $w$ value as $H \left(z \right) \sqrt{\langle w^{2} \left( z \right ) \rangle}_{h}$. 

One of the first trends that can be pointed out is the fact that convective mixing seems to be consistent across every sample in a given effective temperature $T_{\mathrm{eff}}$, independent of rotation period $P_{\mathrm{rot}}$ of a given case. This can be seen by observing each row separately. This matches the design in which each atmosphere is forced from below, as outlined in Section \ref{sec:thermal_forcing_of_the_atmosphere}. The thermal forcing is governed by the amplitude of the imposed thermal perturbations, $T_{\mathrm{amp}}$, which is fixed by the effective temperature (see Section \ref{sec:thermal_forcing_of_the_atmosphere}) and does not vary with rotation period. 

Further trends stand out across our sample, namely all of the $T_{\mathrm{eff}}=400~K$ cases present a uniform, deep, convective column of $K_{zz} = 10^{5}~\rm m^{2}~s^{-1}$ extending from the base of the atmosphere to the $p=1~ \rm bar$ level, tapering off linearly due to convective overshoots, arriving at the minimum level $K_{zz, \mathrm{min}} = 10~\rm m^{2}~s^{-1}$ around $p \approx 0.1~ \rm bar$. For the $T_{\mathrm{eff}}=500~\rm K$ and $T_{\mathrm{eff}}=600~\rm K$ cases, the convective mixing structure appears to be split into two regions, one extending from the base of the pressure domain to about $p=40~\rm bar$, followed by another disjoint convective zone that sits between  $1~ \mathrm{bar} \leq p \leq 10~\rm bar$ for $T_{\mathrm{eff}}=500~K$ and $1~ \mathrm{bar} \leq p \leq 3~\rm bar$ for $T_{\mathrm{eff}} = 600~\rm K$. In a similar study, \citet{lefevre_2022_convection_BD} used a convection‐resolving model and found that their $K_{zz}$ estimates peak at $\approx 10^{6}~m^2~s^{-1}$ within convective zones but drop by several orders of magnitude above them. They caution that these very small post-convective values are not appropriate for GCMs given the lack of wind shear. Our overshoot parameter ($\beta= 2.2$) follows this suggestion and produces a more gradual decline of $K_{zz}$, so the profiles shown here should be read as an upper limit on mixing.

While the individual structure of the advective vertical velocity changes across our sample, all of them fall in a range of $10^{-1} \text{--} 10~\rm m^{2}~s^{-1}$, indicating a regime characterized by weak advective mixing. This is in line with our expectations from these substellar atmospheres, where convective mixing is expected to dominate. In all of our examples, the RCB falls within the $p=1~\text{--}~5~ \rm bar$ level and sits below the pressure level of the photosphere, which is at about $p_{\mathrm{photosphere}} \approx 0.7~\rm bar$ for all the presented runs. 

\subsection{Cloud properties}

\begin{figure*}
    \centering
    
    \begin{subfigure}[b]{\textwidth}
        \centering
        \includegraphics[width=\textwidth]{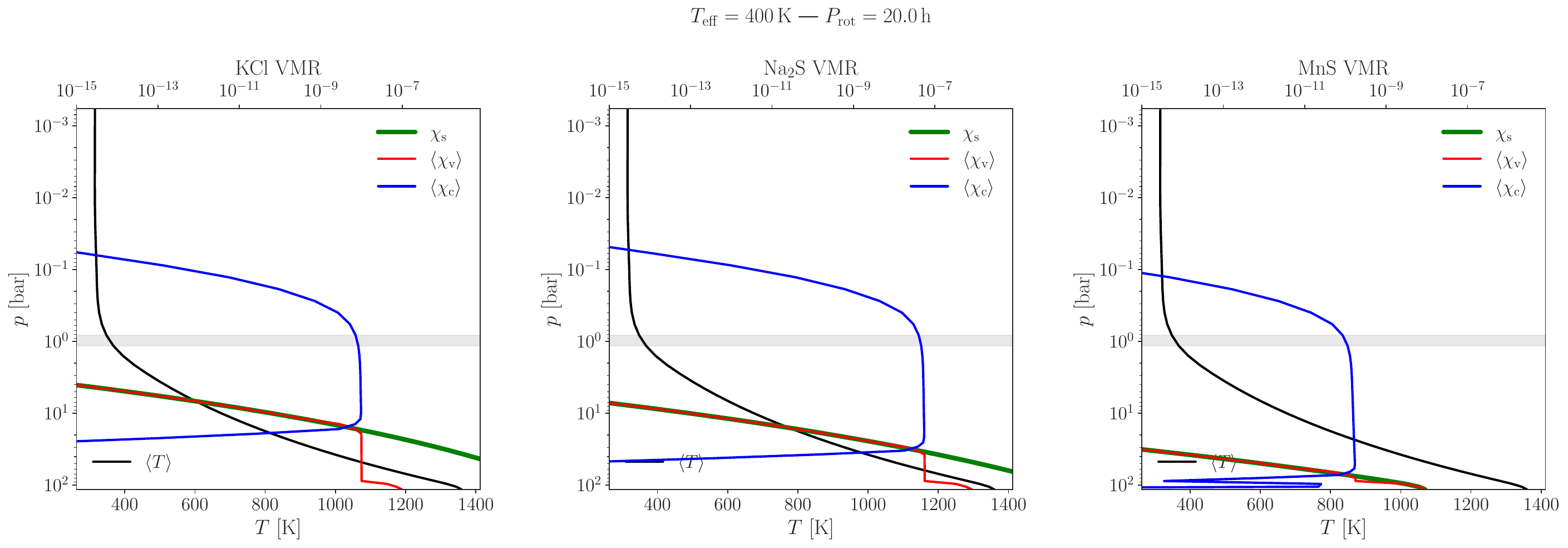}
        \caption{$T_{\mathrm{eff}} = 400\,\text{K}$}
        \label{subfig:vmr_400K}
    \end{subfigure}
    \vspace{0.4em}

    \begin{subfigure}[b]{\textwidth}
        \centering
        \includegraphics[width=\textwidth]{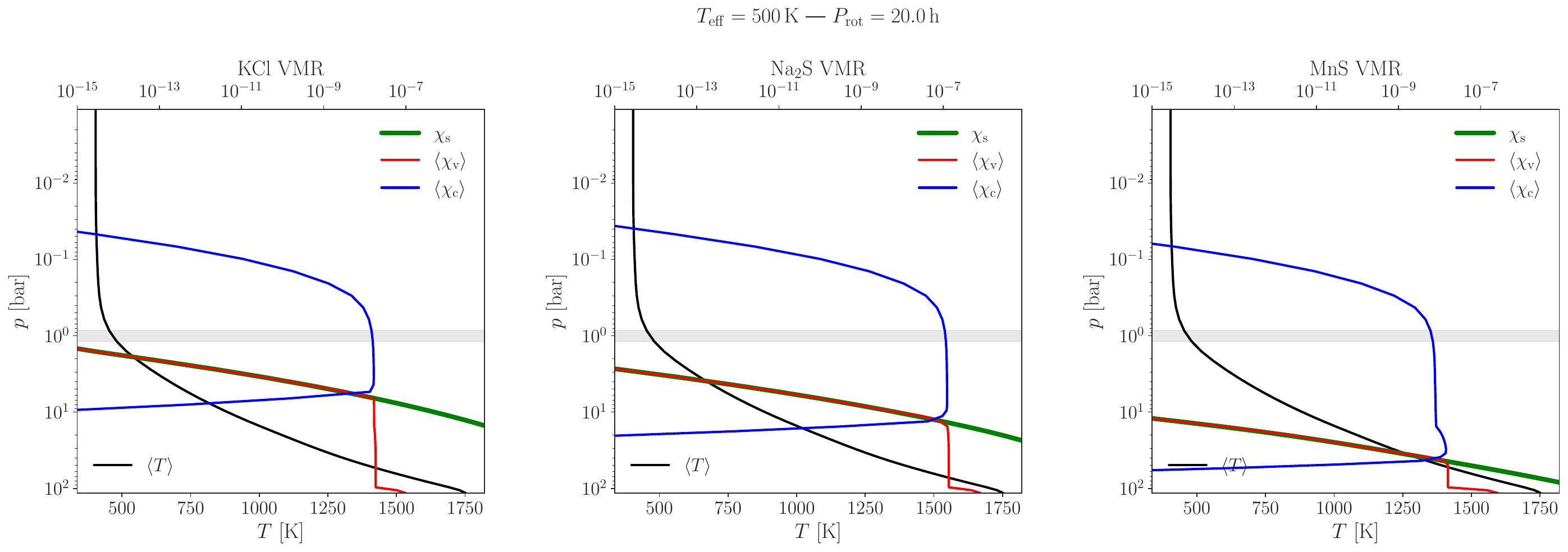}
        \caption{$T_{\mathrm{eff}} = 500\,\text{K}$}
        \label{subfig:vmr_500K}
    \end{subfigure}
    \vspace{0.4em}

    \begin{subfigure}[b]{\textwidth}
        \centering
        \includegraphics[width=\textwidth]{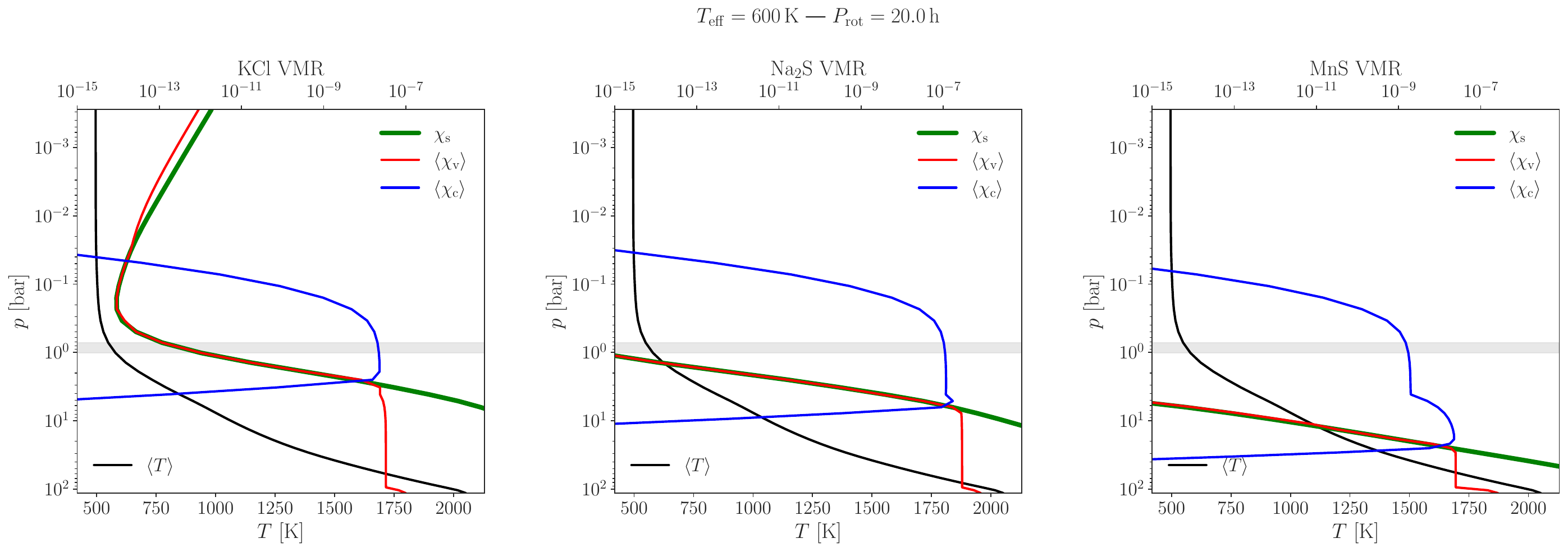}
        \caption{$T_{\mathrm{eff}} = 600\,\text{K}$}
        \label{subfig:vmr_600K}
    \end{subfigure}

    \caption{Globally averaged vertical profiles of cloud condensate (blue) and vapour (red) VMRs, together with the saturation vapour VMR (green), for three cloud species (columns, left to right: $\mathrm{KCl}$, $\mathrm{Na_{2}S}$, $\mathrm{MnS}$) and three effective temperatures (rows, top to bottom: $400 \text{--}600$ K). Each panel shows results at a simulation time of $t_{\mathrm{run}} = 1000\,\text{d}$, averaged over the final $10$ days, for a fixed rotation period of $P_{\mathrm{rot}} = 20\,\text{h}$. The horizontal gray shaded area marks the photosphere for the single-band thermal emission.}
    \label{fig:cloud_vmr_grid_20h}
\end{figure*}

Figure \ref{fig:cloud_vmr_grid_20h} shows, for each case, the thermal band photosphere, the globally averaged VMRs of vapor and condensate for every cloud species together with the equilibrium saturation VMR $\chi_{s}$. The $T\text{--}p$ profile is superimposed to allow direct comparison with the vertical-mixing plots in Figure \ref{fig:Kzz_composite}. Different rows correspond to the models with $T_{\mathrm{eff}} = 400$, $500$, and $600 ~\rm K$ and a fixed rotation period of $P_{\mathrm{rot}} = 20$ h. Although different rotation periods, for the same effective temperature, result in varying horizontal structures within a given layer, the globally-averaged vertical profile of clouds and their associated number density distributions remain nearly identical. That is why a single rotation period, $P_{\mathrm{rot}} = 20~\rm h$ in Figure \ref{fig:cloud_vmr_grid_20h}, suffices to outline the main behaviour. In every panel, the vapor VMR curve, shown in red, follows the saturation curve in green up to the deep-reservoir value listed as $\chi_{\mathrm{deep}}$ in Table \ref{tab:cloud_numerics}, showing that the simulated atmospheres stay saturated with vapor throughout. The condensate VMR curve, plotted in blue, marks the altitudes where cloud condensates settle, forming a cloud layer. As discussed in Section \ref{sec:gravitational_settling_of_clouds}, the vertical cloud profile results from a balance among convective mixing, gravitational settling, and the supply from the deep reservoir. The deep reservoir sets an upper limit, yet whether that limit is reached depends mainly on the strength of vertical mixing, which, in our simulations, is dominated by convection. This link becomes clear when one compares the mixing intensities in Figure \ref{fig:Kzz_composite} with the shapes of the condensate curves in Figure \ref{fig:cloud_vmr_grid_20h}. Gravitational settling confines the cloud layer vertically, and thermal coupling has a similar regulating effect, though by a different route. When cloud thermal coupling is active, part of the internally deposited heat is absorbed and scattered within the cloudy regions, giving a slight warming that evaporates a small fraction of the local condensate and lowers its VMR accordingly.

\begin{figure*}
    \centering
    \includegraphics[width=\textwidth]{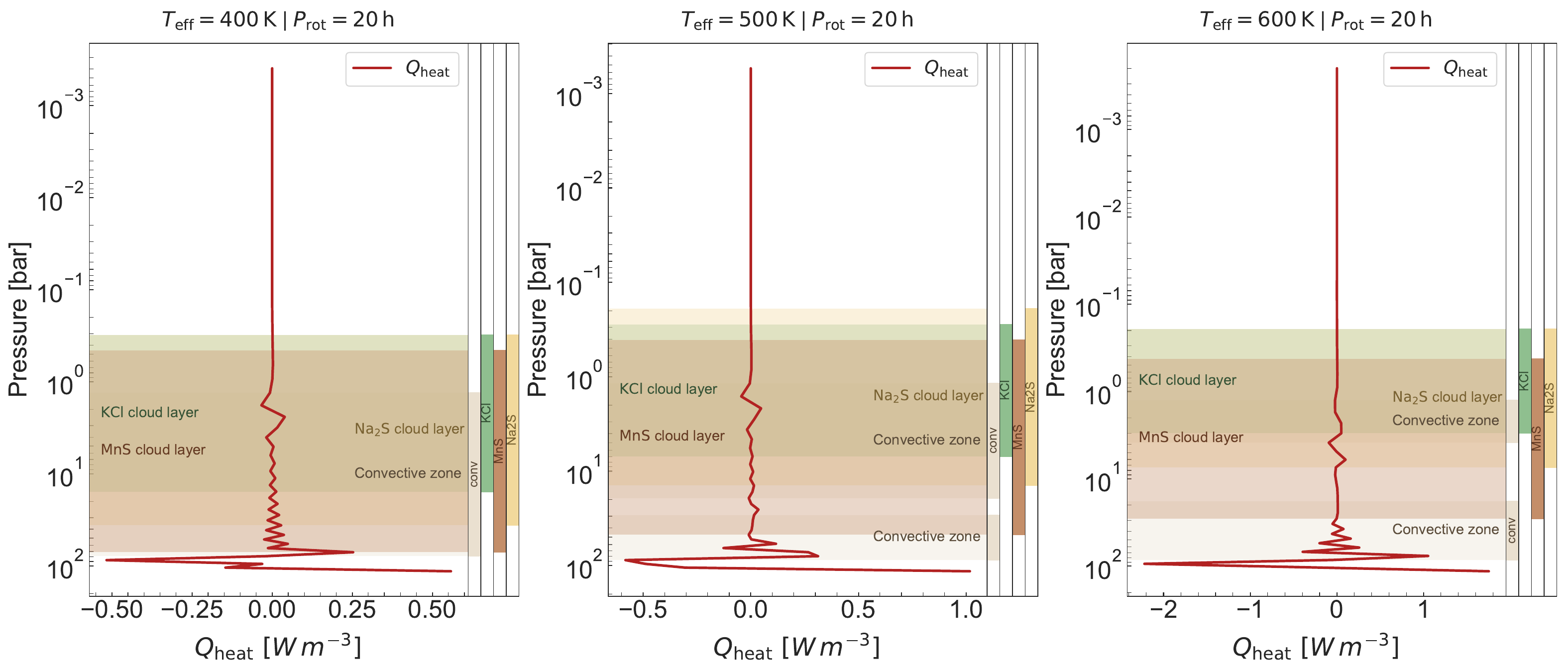}
    \caption{Vertical profiles of diabatic heating rates for the run cases with $T_{\mathrm{eff}} = \left[400, 500, 600 \right]~\text{K}~;~ P_{\mathrm{rot}}=20~\text{h}$ at $t_{\mathrm{run}} = 1000~\rm d$. Overlaid are convective zones and cloud condensate layers, intended to give an overview of the physical processes that contribute to the radiative balance of the atmospheres under study. The smaller panels on the right-hand side shows the extent of each overplotted layer.}
    \label{fig:qheat_profs}
\end{figure*}

For $T_{\mathrm{eff}} = 400~\rm  K$, $\mathrm{KCl}$ and $\mathrm{Na_{2}S}$ clouds appear near $p = 10 ~\rm bar$ and extend upward to about $p = 0.3$ bar, to the level of the brown dwarf's photosphere. $\mathrm{MnS}$ clouds occupy a similar pressure but extend deeper and have significantly weaker abundances. At this effective temperature, the saturation curve lies below the deep-reservoir mixing ratio even at the lower boundary, so the vapor mixed upward becomes supersaturated immediately. Condensate, therefore, collects at the base of the atmosphere, where our model removes it, and its mixing ratio can rise only as far as convective transport can redistribute it. For the $T_{\mathrm{eff}} = 500$ K and $600$ K models, the behavior of clouds is qualitatively similar, though the extent of the cloud layers becomes narrower as the effective temperature increases. While the cloud layers show some overlap with the photosphere level, their optical depths peak around the base of the illustrated condensate curve and remain well below unity ($\tau_{\mathrm{cloud}} \ll 1$) across the whole atmosphere.

Figure \ref{fig:qheat_profs} shows the globally averaged vertical profiles of diabatic heating rate produced by the dynamical core. In each panel, we plot the volumetric heating rate $Q_{\mathrm{heat}}\,[\mathrm{W\,m^{-3}}]$ along with the cloud layers where condensate mixing ratios peak, as indicated in Figure \ref{fig:cloud_vmr_grid_20h}, and the convective zones identified by a positive convective heat flux $F_{\mathrm{conv}} \textgreater ~0$ from the MLT routine. Convective overshoot enhances the mixing of cloud particles but does not contribute directly to $Q_{\mathrm{heat}}$. The $Q_{\mathrm{heat}}$ values shown here are the source terms that enter the model’s energy equation and are therefore expressed per unit volume. They can be converted to a temperature tendency through $dT/dt = Q_{\mathrm{heat}}/(\rho c_{p})$, resulting in peak magnitudes comparable to the thermal perturbations imposed at $p = 10~ \mathrm{bar}$. In the statistically steady state, $Q_{\mathrm{heat}}$ departs from zero primarily in the deep layers where internal forcing and convection act, while above the thermal-band photosphere $(p \approx 0.7~\mathrm{bar})$ it remains close to zero in the global mean. Although cloud layers extend above the convective zones, they remain optically thin in our gray thermal band and do not coincide with comparable enhancements in $Q_{\mathrm{heat}}$ in these profiles. It should be noted that the cloud layers that are shown in Figure \ref{fig:qheat_profs}, do not all correspond to the same VMR, they simply indicate where cloud condensates settle. 

\subsection{Flow regime}
\label{subsection:flow_regime}
Figure \ref{fig:hammond23plots} shows the thermal Rossby number $Ro_{\mathrm{T}}$ plotted against the nondimensional radiative timescale $\hat{\tau}_{\mathrm{rad}}$, following the conventions of \citet{hammond_2023_shallow_water}. The axis ranges match that study, and the gray line marks the boundary separating two of their flow regimes, the upper one with small vortices and an equatorial jet, the lower dominated by radiative forcing and lacking jets (see their Figure 8). Each symbol apart from the red star represents one of the GCM simulations from this study, with $Ro_{\mathrm{T}}$ and $\hat{\tau}_{\mathrm{rad}}$ calculated from our model outputs using the definitions in Section \ref{sec:characteristic_flow_quantities} (see Eq. \ref{eq:geopotential} $\&$ \ref{eq:normalized_tau_rad}). We also include an additional GCM run with $T_{\mathrm{eff}} = 600~\mathrm{K},P_{\mathrm{rot}} = 20\mathrm{h},~\log g = 3.5$, designed to test whether reduced gravity enhances variability. This case is plotted in a lighter green to distinguish it from the standard $T_{\mathrm{eff}} = 600~\mathrm{K}~;~ P_{\mathrm{rot}} = 20~\rm h$ run and its variability implications are discussed in Section~\ref{sec:variability}. All of our simulated cases lie below the plotted regime line, inside the radiative-forcing-dominated regime, and the plotted values also reveal an even more rotation-dominated regime than the theoretical estimates in Figure \ref{fig:expected_regimes} had implied. 

All points belonging to the same effective temperature (same colour) fall on a straight line with a negative slope. This is a simple result of the normalisations adopted in Section \ref{sec:characteristic_flow_quantities} (also seen in Figure \ref{fig:expected_regimes}), as the rotation period increases (i.e., the angular velocity $\Omega$ decreases), the thermal Rossby number $Ro_\mathrm{T}\propto \Omega^{-2}$ grows, whereas the dimensionless radiative timescale $\hat{\tau}_{\mathrm{rad}}\propto \Omega$ shrinks. This implies, for a constant temperature $T$,  $Ro_{\mathrm{T}} \propto \hat{\tau}_{\mathrm{rad}}^{-2}$  and our results reproduce this scaling. Conceptually, small $Ro_\mathrm{T}$ and large $\hat{\tau}_{\mathrm{rad}}$ signal a flow that is strongly constrained by Coriolis forces and has ample time (in units of a rotation) to redistribute heat before radiation can act. The point corresponding to the \citet{lee_ohno_2025} study falls within our sample, albeit at a different point than our $(T_{\mathrm{eff}} = 400~\mathrm{K}~;~P_{\mathrm{rot}} = 10~ \rm h)$ case. We would like to note that the chosen pressure level corresponding to the photosphere affects the radiative timescale expression significantly, although not the thermal Rossby number. A higher photosphere (lower pressure) would push all of the plotted points to the left and account for the difference, since most of the discrepancy is on the x-axis.

Finally, we would like to connect these results to the ones presented in Figure \ref{fig:expected_regimes}. In the absence of the vertical profiles calculated by our GCM, we generated the values in Figure \ref{fig:expected_regimes} using a thermal wind approximation, estimating the thermal Rossby number through thermal wind velocities. In that scenario, the angle of the slope is a function of the magnitude of horizontal winds in the atmosphere, acting as a proxy for dynamical activity. In contrast, the GCM analysis evaluates the geopotential directly (Eq.~\ref{eq:geopotential}) and reports $Ro_{\mathrm{T}}$. If we were to plot, instead, the standard Rossby number from the GCM winds ($U \sim 1\text{--}10~\mathrm{m~s^{-1}}$), the points would lie at $Ro \sim 10^{-7} \text{--} 10^{-6}$ and the corresponding slopes of points sharing the same effective temperature would be smaller.

\section{Discussion}
\label{sec:discussion}
\subsection{Dynamical activity}
Across our suite of Y dwarf simulations, the results place all cases firmly in the radiative-forcing-dominated regime of the \citet{hammond_2023_shallow_water} study. While this is consistent with the weak dynamical signatures seen in the effective radiating temperature maps and with the near-uniform horizontal temperatures in the $T\text{--}p$ profiles, compared to our theoretical estimations in Section \ref{sec:characteristic_flow_quantities}, we end up producing less dynamical activity in the modelled atmospheres. Within each $T_{\mathrm{eff}}$ sequence, decreasing $P_{\mathrm{rot}}$ produces more pronounced structures but does not organize flow into coherent jets. Flux variations, while very small, seem to be concentrated around the polar regions with a more uniform equatorial region. Over the course of our simulations, a quasi-oscillatory equatorial pattern develops (visible in the $T_{\mathrm{eff}} = 400~\mathrm{K},~500~\mathrm{K}~;~P_{\mathrm{rot}} = 10~\mathrm{h},~20~\mathrm{h}$ cases), persists for a few hundred Earth days, and then transitions into a uniform zonal band structure. The associated flux variations, likely reflecting the imposed internal thermal perturbations, remain negligible. The general lack of strong circulation in our results can be attributed to the weak wave activity. All models exhibit characteristics of mid-to-high-latitude Rossby waves, produced by rapid rotation interacting with the latitudinal gradient of the Coriolis force. The primary indicator of the presence of Rossby waves is the hemispherically antisymmetric tilts, visible in variations of the photospheric temperature (Figure \ref{fig:olr_grid}), seeded by our internal thermal perturbations. These tilted features can be observed in higher latitudes, indicating westward traveling waves, matching the characteristics of Rossby waves. In principle, the interaction of Rossby waves with the mean flow could excite and maintain zonal jets, as suggested by \citet{tan_2021a}, but the lack of such features suggests that the strength of the excited waves compared to radiative processes is not efficient enough to drive such activity.

\begin{figure*}[t!]
    \centering
    \includegraphics[width=0.9\textwidth]{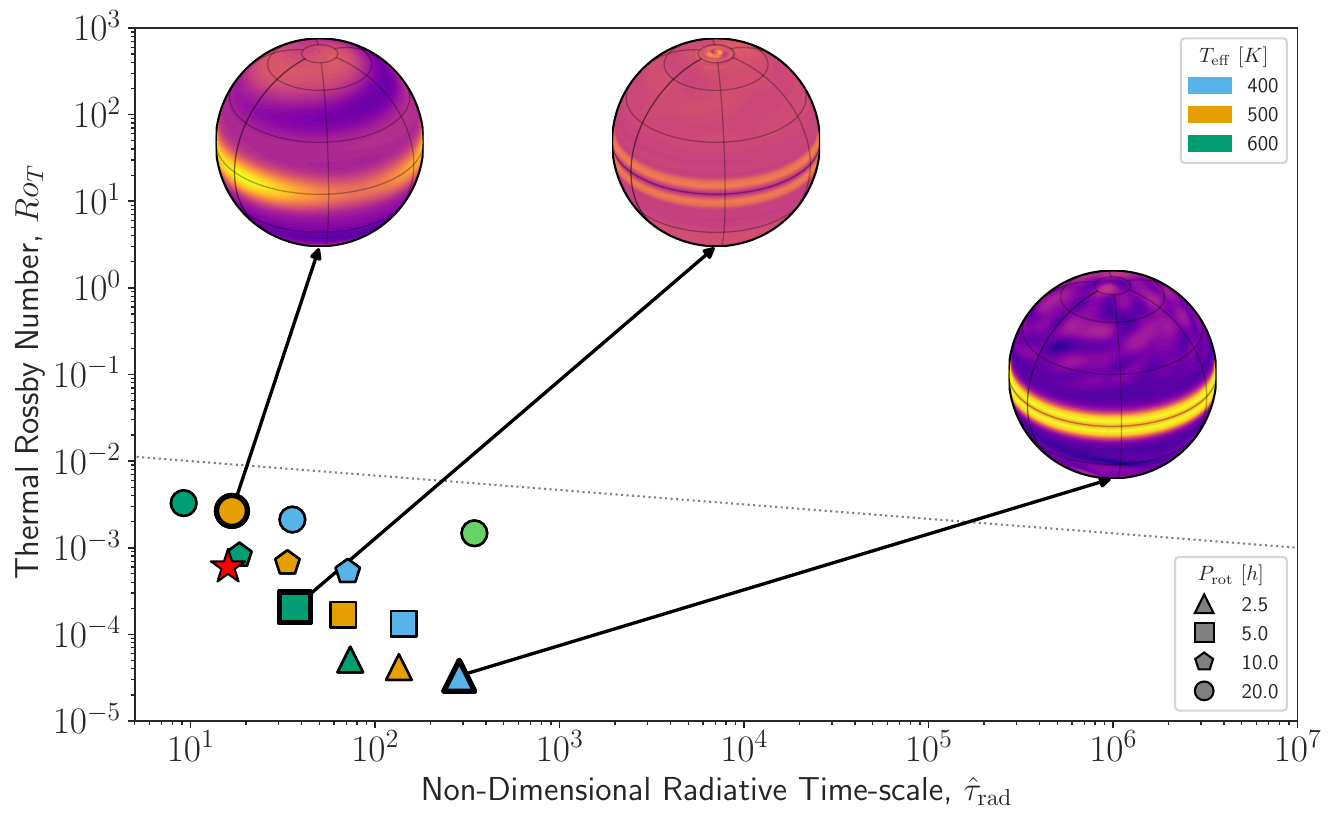}
    \caption{Thermal Rossby number $Ro_{\mathrm{T}}$ plotted against the non-dimensional radiative time-scale $\hat{\tau}_{\rm rad}$ for the complete sample of planets discussed in this study. The plotted quantities are calculated according to the definitions given in Sec. \ref{sec:characteristic_flow_quantities} and at the approximate photosphere corresponding to $p =  0.7~\rm bar$. Chosen examples are highlighted together with their OLR profiles overplotted in the figure. The dashed lines correspond to a boundary between circulation regimes identified by \citet{hammond_2023_shallow_water}. The red star is included as a reference point, the same as in Figure \ref{fig:expected_regimes}, corresponding to a run case of ($T_{\mathrm{eff}} = 400~\mathrm{K}~;~ P_{\mathrm{rot}} = 10~\rm h$; $\log g = 4.46$) from a recent GCM study.}
    \label{fig:hammond23plots}
\end{figure*}

\subsection{Cloud radiative feedback}

Our vertical mixing analysis indicates that convection dominates transport over the weak advective contribution throughout the photosphere and below. This dominance is reflected in the cloud distributions (Figure \ref{fig:cloud_vmr_grid_20h}), condensate layers form where the saturation curve and deep reservoirs permit, but their amplitudes and vertical extents are shaped primarily by convective mixing and gravitational settling. If the convective diffusion terms (in equations \ref{eq:vapour_tracer} $\&$ \ref{eq:condensate_tracer}), whose strength is determined by the $K_{zz}$ values from our mixing-length scheme, is omitted and the dynamical core alone advects cloud tracers upward, the condensate VMR peaks at roughly the same pressures (Figure \ref{fig:cloud_vmr_conv_vs_adv}) but attains values several orders of magnitude lower than those plotted in Figure \ref{fig:cloud_vmr_grid_20h} and is more vertically confined. This implies that, despite clouds not having a strong effect on our presented results, their coupling to convective mixing is crucial for the resulting dynamical structure, if clouds are expected to contribute significantly.

Our results show that clouds are replenished efficiently but do not seed larger-scale dynamical structures. Thermal coupling within cloudy layers slightly warms the local environment, yet these effects remain secondary in the global energy balance. Previous studies \citep{tan_2021b, lee_2024_BD_clouds} had indicated that cloud heating could drive a self-sustaining evaporate–cool–recondense cycle and act as a source of dynamical activity. In our models, this effect is too weak to significantly influence the circulation. Notably, \citet{tan_2021b} explicitly state that their simulation setup was intended to study a cloud-driven dynamical scenario rather than to represent a realistic parameter set. 

We note that our results are obtained with a gray radiative transfer scheme, an approach that is widely used in brown dwarf and exoplanet GCMs \citep[e.g.][]{tan_2021b, malsky_2024, kennedy_2025}. This method does not explicitly resolve spectral windows that sample deep, cloud-bearing layers and may therefore underestimate cloud radiative effects in those bands. Given the low condensate abundances and cloud optical depths in our simulations, we expect this bias to be modest for the parameter space explored here. This implies that our inference of weak cloud feedback should be viewed as conservative rather than invalidated by the gray approximation. Moreover, building on \citet{lee_2024_BD_clouds} with a more realistic cloud model, \citet{lee_ohno_2025} apply the framework to a Y dwarf and obtain the same dynamical regime as our models, as noted in our results. Taken together, our results do not contradict previous studies; rather, they extend that framework to Y dwarf conditions.

In all models, our convective regions coincide with the pressure ranges where the base of the cloud layers forms (Figure \ref{fig:qheat_profs}) and where we impose thermal perturbations. This likely explains the insignificant cloud thermal feedback, as convective heat fluxes dominate the energy transport in these layers. A further possibility is that cloud number densities remain too low to yield a significant contribution to the Rosseland-mean opacity. This interpretation is consistent with Figure \ref{fig:qheat_profs}, where cloud layers extend above the convective zones but the diabatic heating rates show no corresponding increase.

\subsection{Variability}
\label{sec:variability}
Together, these results suggest clear directions for a more dynamically active test case. Following the scalings in our methodology, the most promising configuration would combine a slow rotation rate with the hotter end of our sample, and lower surface gravity (e.g., $\log g=3.5$). Such a setup should result in longer radiative timescales, while the choice of higher effective temperatures more strongly seeds variability, thereby allowing dynamical activity to organize into larger-scale structures and enhance variability. In Figure~\ref{fig:hammond23plots}, we explore such a case and observe that it shifts in the intended direction in the $\left(Ro_{\mathrm{T}} \text{--} \hat{\tau}_{\mathrm{rad}}\right)$ space. While this indicates a more dynamically active regime (further discussed in the Appendix), we still remain in the radiative-forcing-dominated regime. Overall, this extension, while not exhaustive, supports the trend indicated by our main GCM runs: within the physical processes considered here, Y-dwarf variability naturally lies in this regime.

Another direction to enhance variability is to explore alternative cloud-particle size distributions, which could strengthen cloud–radiation coupling. Finally, incorporating disequilibrium chemistry, in future works, will be important to assess whether compositional feedbacks alter the thermal structure and the variability observed here.

\section{Conclusions}
We have presented a curated sample of twelve GCM runs spanning the Y dwarf regime ($400~\mathrm{K} \leq T_{\mathrm{eff}} \leq 600~\mathrm{K}$), to map the dynamical behavior of these atmospheres across a first, parameter-space-wide study. The model setup is observationally motivated and couples a parametrized convection scheme with a simple saturation-adjustment cloud model, enabling a self-consistent assessment of how convective mixing and cloud formation interact to shape the flow, an effect studied in previous studies \citep[e.g.,][]{tan_2021b} on brown dwarf variability. 

Our results can be summarized as follows: Salt and sulfide species ($\mathrm{KCl}, \mathrm{Na_{2}S}, \mathrm{MnS}$) form patchy (Appendix \ref{sec:appendix_c}) and optically thin cloud layers close to the photosphere, but their radiative feedback, and thus, effect on the $T\text{--}p$ profile is minimal. Across the grid, the circulation remains in the radiative-forcing-dominated regime \citep{hammond_2023_shallow_water}, with small-scale, off-equatorial features that do not organize into jets. The $T\text{--}p$ profiles and thermal emission equilibrate within a few hundred days, and top-of-atmosphere temperature variations remain small, with the observed temporal evolution primarily reflecting the imposed thermal perturbations rather than strong feedbacks among included processes. In the presence of the sluggish dynamical activity that we observe, variability is primarily controlled by rotational modulation from long-lived vortices. Overall, these findings align with the theoretical predictions of \citet{hammond_2023_shallow_water}, in that no large-scale dynamical features are present in this circulation regime, with the atmospheres primarily controlled by radiative processes.

Our conclusions are also in line with the recent results of \citet{lee_ohno_2025}, despite the inclusion of more cloud species and a wider parameter range coverage, supporting the view that models of Y dwarf atmospheres produce little dynamical activity in general. In particular, cloud–radiative effects in this regime appear insufficient to substantially alter the large-scale dynamical state or to induce new dynamical features. At the same time, our single-band, gray radiative transfer with Rosseland-mean opacities does not account for spectral windows that could probe deep, cloud-rich layers and may therefore underestimate cloud radiative feedback in those bands. In our simulations, however, the condensate mixing ratios keep the clouds optically thin near the photosphere, so their radiative influence remains small. Our conclusions should thus be interpreted as applying to atmospheres with relatively low cloud optical depths and do not preclude stronger feedback in cases with a larger deep vapor reservoir.

These results suggest that the dominant mechanisms behind Y dwarf variability remain uncertain. Although our predicted inhomogeneities are slight, the sensitivity of \textit{JWST} means that contrasts of a few Kelvin can affect the spectra and should be detectable. A more sophisticated modelling approach, with higher resolution dynamics, improved cloud microphysics and radiative coupling, and disequilibrium chemistry, will help clarify the picture, but progress depends on future \textit{JWST} observations analogous to recent variability studies \citep[e.g.,][]{biller_2024, chen_2025} of L–T dwarfs.

\bibliographystyle{aa.bst}
\bibliography{bib}

\begin{appendix}
\section{Adaptation of the tracer equations}
\label{sec:appendix_adaptation_of_the_tracer_equations}
Our 2-species tracer-based relaxation model can be generally written as,
\begin{equation}
    \frac{Dq}{Dt} = \hat{\mathcal{S}},
\end{equation}
where $\frac{D}{Dt} = \frac{\partial}{\partial t} + \mathbf{v} \cdot \nabla$ is the material derivative, $q$ is the MMR of the vapor or the condensate, $\mathbf{v}$ the wind field and $\hat{\mathcal{S}}$ are the related source terms. We would like to rewrite these equations as a standard continuity equation with a source term:
\begin{equation}
    \frac{\partial \rho q}{\partial t} + \nabla \cdot \left( \rho \mathbf{v} q \right) = \mathcal{S}.
\end{equation}
In this form, we can pass the tracer MMRs directly to the dynamical core (as explained in Section \ref{sec:modelling_of_clouds}), since the entropy equation solved by the dynamical core has the same form and is readily discretized on the icosahedral grid. The source terms can then be integrated in the physics modules and passed to the dynamical solver as correction terms, for it to advect tracer concentrations accordingly. The tracer equations for the vapor and the condensate species are given by:
\begin{align}
    \frac{Dq_{v}}{dt} &= \left(1 - s \right) \frac{\text{min}\left(q_{s} - q_{v}, q_{c}\right)}{\tau_{c}} - s \frac{\left(q_{v} - q_{s}\right)}{\tau_{c}} - \frac{q_{v} - q_{\text{deep}}}{\tau_{\text{deep}}}, \label{eq:tracer_equations_v} \\
    \frac{Dq_{c}}{dt} &= s \frac{\left(q_{v} - q_{s} \right)}{\tau_{c}} - \left(1 - s \right) \frac{\text{min}\left(q_{s} - q_{v}, q_{c}\right)}{\tau_{c}}, \label{eq:tracer_equations_c}
\end{align}
Expanding the material derivative terms, we can write,
\begin{align}
    \frac{D}{Dt} \left( \rho q_{c} \right) &= \frac{\partial \left( \rho q_{c} \right)}{\partial t} + \mathbf{v} \cdot \nabla \left( \rho q_{c} \right), \\
    &= \frac{\partial \left( \rho q_{c} \right)}{\partial t} +  \nabla \cdot \left( \rho q_c \mathbf{v} \right) - \rho q_c \left( \nabla \cdot \mathbf{v} \right), \\
    \implies \frac{D}{Dt} (\rho q_c) &= \frac{\partial (\rho q_c)}{\partial t} + \nabla \cdot (\rho q_c \mathbf{v}) + q_c \frac{D \rho}{Dt}, \\
    \implies \rho \frac{D q_c}{Dt} &= \frac{\partial (\rho q_c)}{\partial t} + \nabla \cdot (\rho q_c \mathbf{v}),
\end{align}
whereby we have used the vector calculus identity,
\begin{equation}
    \nabla \cdot (\phi \mathbf{v}) = \phi (\nabla \cdot \mathbf{v}) + \mathbf{v} \cdot \nabla \phi,
\end{equation}
and the mass continuity equation,
\begin{equation}
    \frac{\partial \rho}{\partial t} + \nabla \cdot (\rho \mathbf{v}) = 0.
\end{equation}
We can rewrite the expanded material derivative in the same form as the above-mentioned tracer equations \ref{eq:tracer_equations_v} $\&$ \ref{eq:tracer_equations_c} , yielding,

\begin{align}
    \frac{\partial (\rho q_v)}{\partial t}
      + \nabla\!\cdot (\rho \mathbf{v} q_v)
      &= \rho \Biggl[
          (1-s)\frac{\min(q_s-q_v,\,q_c)}{\tau_c}
          - s\frac{q_v-q_s}{\tau_c} \notag\\
      &\qquad
          - \frac{q_v-q_{\text{deep}}}{\tau_{\text{deep}}}
         \Biggr], \notag\\[4pt]
    \frac{\partial (\rho q_c)}{\partial t}
      + \nabla\!\cdot (\rho \mathbf{v} q_c)
      &= \rho \Biggl[
          s\frac{q_v-q_s}{\tau_c}
          - (1-s)\frac{\min(q_s-q_v,\,q_c)}{\tau_c}
         \Biggr]                  
\end{align}
\section{Higher variability case results}
\begin{figure}
    \centering
    \includegraphics[width=0.75\columnwidth]{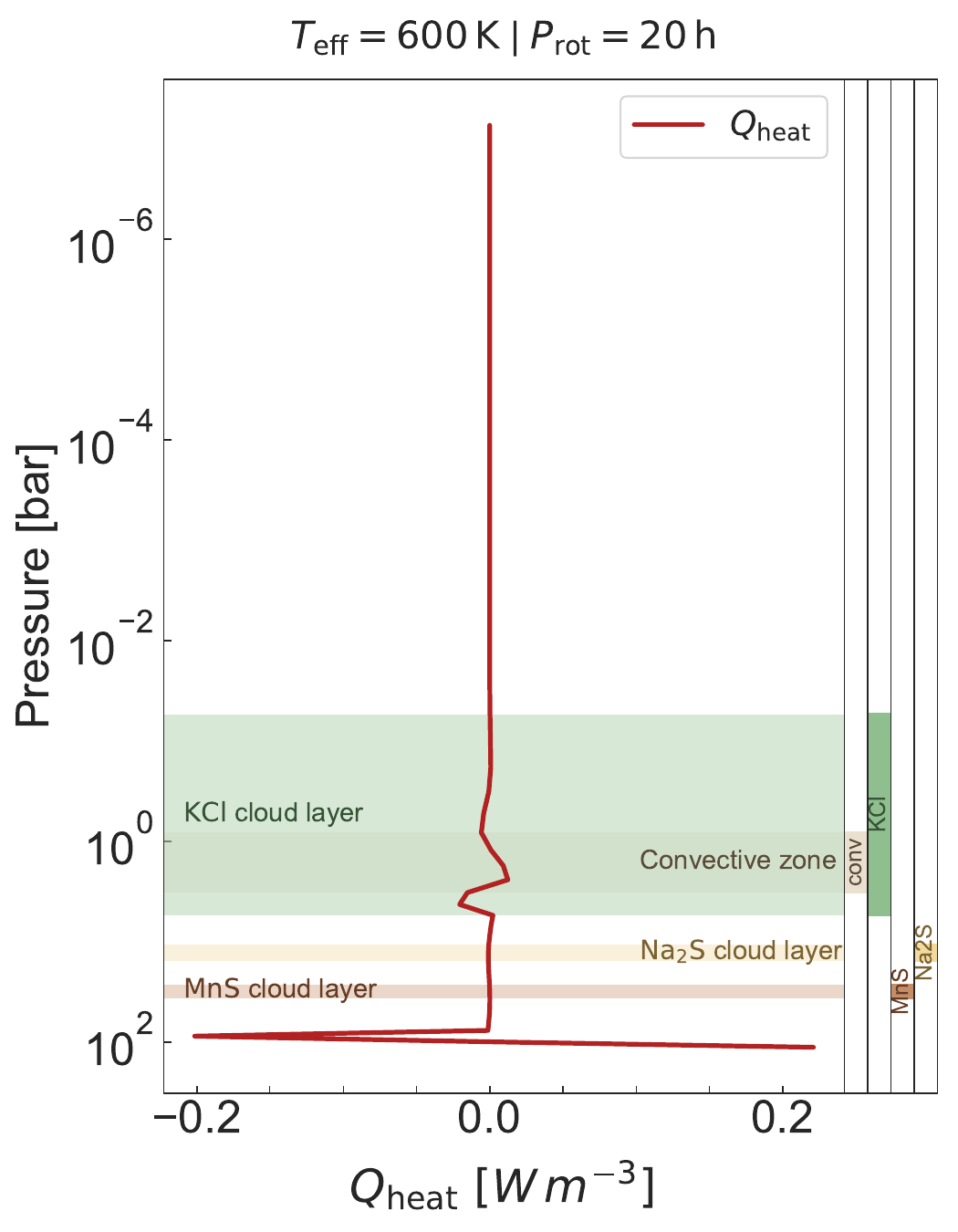}
\caption{Vertical profiles of diabatic heating rates for the run case $T_{\mathrm{eff}} = 600 ~\mathrm{K}~;~ P_{\mathrm{rot}}=20~\mathrm{h}~; \log g = 3.5$ at $t_{\mathrm{run}} = 500~\rm d$. Overlaid are convective zones and cloud condensate layers. The smaller panels on the right-hand side shows the extent of each overplotted layer.}
    \label{fig:qheat_lowg}
\end{figure}

Here we present a quick summary of the “higher variability” case from the Discussion Section. The only change in the setup of the model is a lower gravity of $\log g = 3.5$. As discussed in the main text, this increases the radiative timescale and allows dynamical structures to form. In this case an equatorial jet emerges with $u \sim 30~\rm m~s^{-1}$, and the temperature field shows a broad, nearly uniform, and generally warmer (about $0.5~\rm K$) equatorial region with similarly broad, uniform bands at higher latitudes. This supports the trend in Section \ref{sec:results} from small high-latitude vortices to larger zonal band structure focused near the equator.

The convective mixing and the extent of the cloud layers are narrower, as shown in Figure \ref{fig:qheat_lowg}. Together with the more zonally uniform temperature field, this suggests that the increased variability is driven by atmospheric wave activity rather than cloud feedback, which would produce more complex, small scale variations.

\section{Cloud distributions}
\label{sec:appendix_c}

\begin{figure*}
    \centering
    \includegraphics[width=\textwidth]{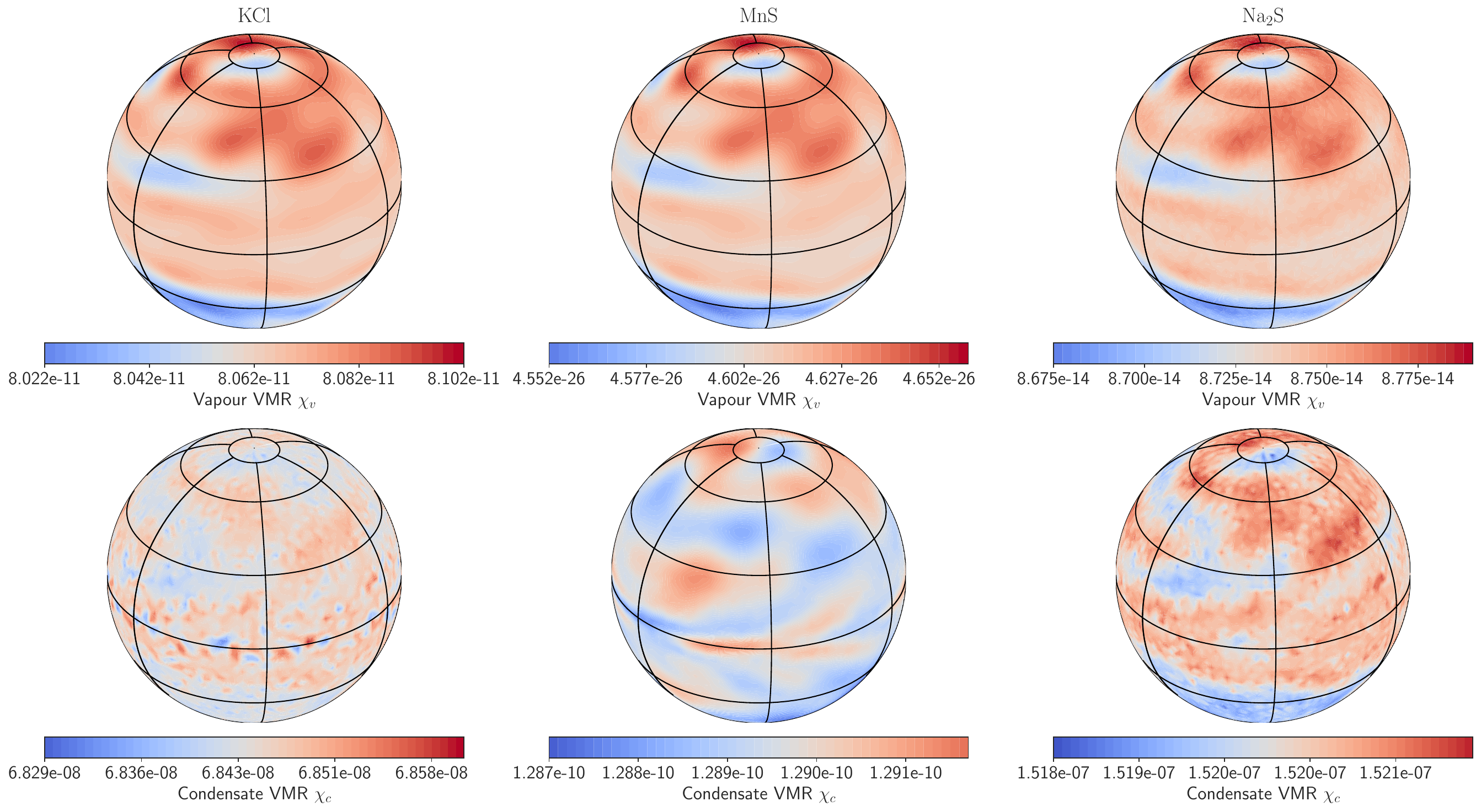}
    \caption{Example cloud distribution plot for $T_{\mathrm{eff}} = 600\,\text{K}~;~ P_{\mathrm{rot}}=20~\text{h}$ from a snapshot at a runtime of $t_{\mathrm{run}} = 500~\rm d$ at $p\approx 1~\rm bar$.}
    \label{fig:cloud_vmr_level}
\end{figure*}

In the parameter space we explore in this study, cloud radiative feedback remains ineffective in driving the atmosphere. For completeness, we report the horizontal cloud distributions at the level where the cloud condensates settle. Figure \ref{fig:cloud_vmr_level} displays the VMRs of condensable cloud vapours and condensates at a pressure of $p \approx 1~\rm bar$ for our run case $T_{\mathrm{eff}} = 600~\rm K ~;~ P_{\mathrm{rot}} = 20~\rm h$. The top row, showing the vapour VMRs, closely follows the underlying temperature field, with amplitudes differing by orders of magnitude due to differences in condensation curves. The cloud condensates also track the temperature structure but show slight variations. Both $\mathrm{KCl}$ and $\mathrm{Na_{2}S}$ clouds display granular structures with higher overall VMRs than $\mathrm{MnS}$, which shows a more continuous structure. This may reflect the higher VMRs and thus number densities of $\mathrm{KCl}$ and $\mathrm{Na_{2}S}$, which strengthen local evaporation, cooling, and recondensation cycles. VMR variations remain below $1\%$, consistent with the overall ineffectiveness of cloud thermal feedback in these simulations. While we present a single case here, the described behavior captures the overall trends in our simulated sample: cloud structures largely mirror the underlying temperature field, with the degree of granularity set by the condensate VMRs at the cloud bases. These results largely indicate a lack of strong feedback mechanisms. The cloud patchiness is dictated by temperature variations due to the long-lived vortices present in the atmosphere.

\begin{figure*}
    \centering
    \includegraphics[width=\textwidth]{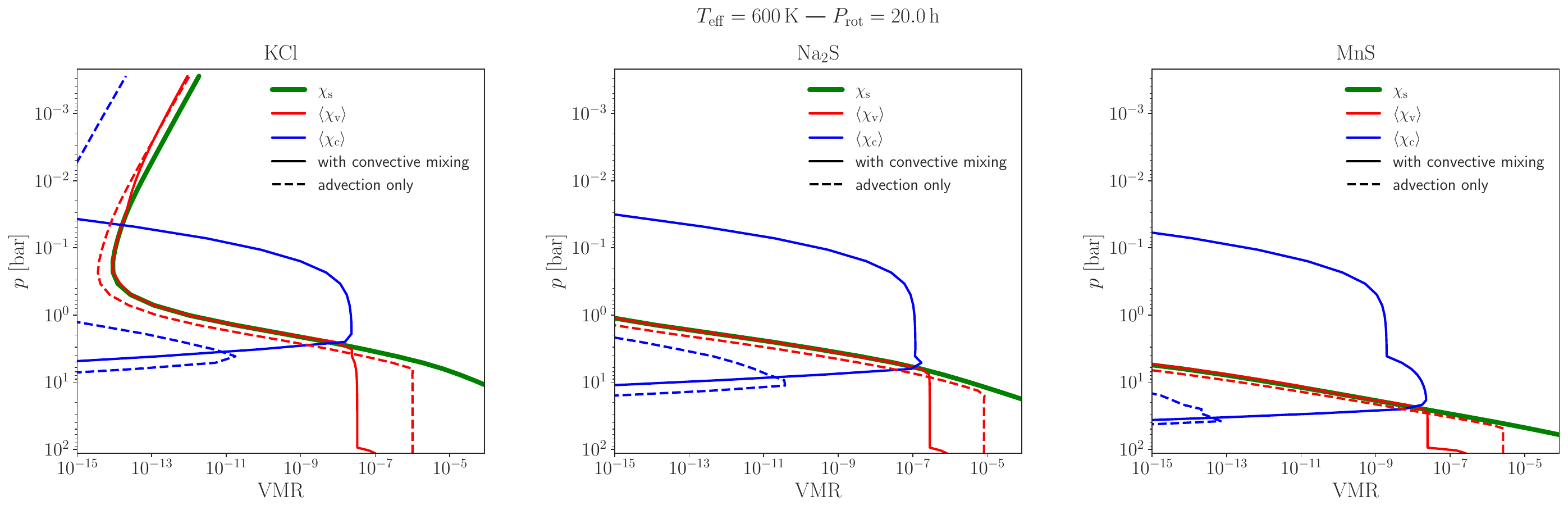}
    \caption{Globally averaged vertical profiles of vapour (red) and condensate (blue) VMRs, together with the saturation vapour VMR (green), for $\mathrm{KCl}$, $\mathrm{Na_{2}S}$, and $\mathrm{MnS}$ (left to right) in the $T_{\mathrm{eff}}=600~\mathrm{K}~;~ P_{\mathrm{rot}}=20~\mathrm{h}$ case at $t_{\mathrm{run}}=1000~\mathrm{d}$ averaged over the final 10 days. Solid curves show the fiducial model including convective mixing; dashed curves show an advection-only variant with convective diffusion omitted.}
    \label{fig:cloud_vmr_conv_vs_adv}
\end{figure*}

Lastly, we use the same run case as an illustrative example to highlight the differences between pure advection, in which the convective diffusion terms in Eqs \ref{eq:vapour_tracer} $\&$ \ref{eq:condensate_tracer} are omitted and tracers are transported only by advection in the dynamical core and our fiducial cases presented in the main body of the paper including convective mixing of the tracers. Figure \ref{fig:cloud_vmr_conv_vs_adv} shows a comparison between these two cases and, other than the omission of the overlaid $T\text{--}p$ profile, is identical to Figure \ref{fig:cloud_vmr_grid_20h} As stated in Section~\ref{sec:discussion}, the condensate layers peak at similar pressures, but their peak VMR values decrease by orders of magnitude when convective mixing is removed, demonstrating that convective mixing dominates the replenishment of condensates for the parameter space explored here.

\end{appendix}

\end{document}